\documentclass[12pt]{article}
\usepackage{mathrsfs,extarrows}
\usepackage{url}
\usepackage{longtable}
\usepackage{caption}
\usepackage[doublespacing]{setspace}
\usepackage[colorlinks=true,citecolor=blue,linkcolor=blue]{hyperref}%
\usepackage{amsthm}
\usepackage{authblk}
\usepackage{subfigure} 
\usepackage{graphicx}
\usepackage{dsfont}
\usepackage{JASA_manu}
\usepackage{array}
\usepackage{amsmath,color}
\usepackage{amssymb,enumerate}
\usepackage{natbib,epsfig,graphicx,rotating}
\usepackage{algorithm,algpseudocode,float}
\usepackage{lipsum}
\usepackage{bm}
\usepackage{color,xcolor}

\makeatletter
\newenvironment{breakablealgorithm}
  {
   \begin{center}
     \refstepcounter{algorithm}
     \hrule height.8pt depth0pt \kern2pt
     \renewcommand{\caption}[2][\relax]{
       {\raggedright\textbf{\ALG@name~\thealgorithm} ##2\par}%
       \ifx\relax##1\relax 
         \addcontentsline{loa}{algorithm}{\protect\numberline{\thealgorithm}##2}%
       \else 
         \addcontentsline{loa}{algorithm}{\protect\numberline{\thealgorithm}##1}%
       \fi
       \kern2pt\hrule\kern2pt
     }
  }{
     \kern2pt\hrule\relax
   \end{center}
  }
\makeatother
\begin{document}
\title{Interpreting and predicting the economy flows: A time-varying parameter global vector autoregressive integrated the machine learning model}
\author[a]{Yukang Jiang}
\author[b,c]{Xueqin Wang}
\author[a]{Zhixi Xiong}
\author[d]{Haisheng Yang}
\author[a,1]{Ting Tian}
\affil[a]{School of Mathematics, Sun Yat-sen University}
\affil[b]{School of Management,University of Science and Technology of China}
\affil[c]{School of Statistics, Capital University of Economics and Business}
\affil[d]{Lingnan college, Sun Yat-sen University}
\begin{footnotetext}[1]
{Corresponding author: tiant55@mail.sysu.edu.cn}
\end{footnotetext}
\date{\today} 

\maketitle
\begin{abstract}
The paper proposes a time-varying parameter global vector autoregressive (TVP-GVAR) framework for predicting and analysing developed region economic variables. We want to provide an easily accessible approach for the economy application settings, where a variety of machine learning models can be incorporated for out-of-sample prediction. The LASSO-type technique for numerically efficient model selection of mean squared errors (MSEs) is selected. We show the convincing in-sample performance of our proposed model in all economic variables and relatively high precision out-of-sample predictions with different-frequency economic inputs. Furthermore, the time-varying orthogonal impulse responses provide novel insights into the connectedness of economic variables at critical time points across developed regions. We also derive the corresponding asymptotic bands (the confidence intervals) for orthogonal impulse responses function under standard assumptions. 
\end{abstract}


\newpage
\section{Introduction}
\indent\indent Global economic integration is inherently essential for country-wise interdependent and country-specific autonomy. Economic activity is once internationally started, and its transmission provides insights into the economic advancements across and among countries. The activity reaction does affect a government and global authorities, which leads to evaluating the identification of the activity effect across countries. Methods for implementing the reaction function for economic activity become vital due to the curse of global-wide aggregation. Global vector autoregressive (GVAR) is a means of simultaneously explicating numerous economies' dynamics \citep{pesaran2004modeling}. In the sense of transmission mechanism for the economic activity, the impulse response function (IRF) is typically employed to depict activity (shock) within cross-sections. Mixed cross-section GVAR (MCS-GVAR) evaluates the interaction between different cross-sectional types \citep{gross2013measuring}. However, it is under the endogenous assumption for the model set. \citet{georgiadis2015examining} introduced exogenous variables in an MCS-GVAR to retain a common economic shock as a cross-section. The author effectively solved the issue of inconsistent timing and frequency of distinct variables and considered the interaction between variables in different locations and the overall regulation approach of the common economic shock. However, regardless of the endogenous and exogenous variables in those models, their weights are time-invariant. Therefore, the impacts of fitting and forecast accuracy are not desirable.

Alternatively, \citet{hauzenberger2021bayesian} considered a time-varying panel vector autoregressive. This model had a relatively high prediction accuracy but did not take into account the mutual impact between different variables in different regions and the overall regulatory of common economic shock. The parameter structure obtained by training in this model cannot be extended to future predictions, therefore, it is impossible to forecast future sequences, and the interpretability is insufficient. We consider modelling time-varying weights for cross-sectional variables. Given the time-varying weights, the bounds of the impulse response could be obtained. In particular, the shock change is more general, with much greater applicability. We apply for the orthogonal IRF (OIRF) \citep{sims1980macroeconomics,kilian2017structural}. Given the OIRF based on our time-varying parameter GVAR (TVP-GVAR) model, we will calculate their bounds using asymptotic distribution \citep{lutkepohl2020constructing,lutkepohl2005new} and examine the significance of the shock. In the framework of TVP-GVAR, we introduce methods for country-specific and global-wide responses that vary over time, consider the economic variables (such as gross domestic product, GDP; and unemployment rate) spreads at multiple time intervals, and integrate machine learning models of the proposed TVP-GVAR model, referred to as TVP-GVAR-ML. 
      
The general form of GVAR model \citep{pesaran2004modeling} is defined as:
\begin{equation}\label{equ:1}
\boldsymbol{x}_{t}=\boldsymbol{b}+\boldsymbol{F}_{1} \boldsymbol{x}_{t-1}+\boldsymbol{\varepsilon}_{t},
\end{equation}
where 
$$\boldsymbol{b}=\boldsymbol{G}_{0}^{-1} \boldsymbol{a},\quad \boldsymbol{F}_{1}=\boldsymbol{G}_{0}^{-1} \boldsymbol{G}_{1},\quad \boldsymbol{\varepsilon}_{t}=\boldsymbol{G}_{0}^{-1} \boldsymbol{u}_{t}.
$$
Originally, $\boldsymbol{G}_{0}$ is a non-singular matrix and not varied over time, and $\boldsymbol{u}_{t}$ is residuals. Thus, parameters $\boldsymbol{b}$ and $\boldsymbol{F}_{1}$ are assumed here as time-invariant parameters of Equation (\ref{equ:1}). 

The setting above allows $\boldsymbol{G}_{0}$ to vary, which enables a more general assessment of macroeconomic regulation. We also establish different machine learning models for our estimators of time-varying weights. As \citet{farrell2021deep} stated, these based on our proposed models are offered as training samples in the various machine learning models to forecast coming trajectories and then put back into our models to obtain economic variables. When the weights are correctly estimated, data-driven learning may be applied to exhibit superior performance in the trajectories of economic variables. Thus, our models will have a two-stage analysis for out-of-sample prediction. We attempt to identify the economic activity response, technical capabilities needed to be user-friendly and interpret the estimates to maximize their values for authorities.   

In Section \ref{sec2} we introduce our models, propose GRF and compute their boundaries. The corresponding analysis of economic data in the United States, Euro area, and Japan are illustrated in Section \ref{sec3}. In the end, section \ref{sec4} discusses the remarks and inference of the proposed model.      

\section{Methodologies}\label{sec2}
\subsection{Data sources}
\indent\indent We consider economic data from Organisation for Economic Cooperation and Development (OECD, \url{https://data.oecd.org/}). The monthly consumer price index (CPI, \url{https://data.oecd.org/price/inflation-cpi.htm}), Harmonised unemployment rate (HUR, \url{https://www.oecd.org/sdd/labour-stats/unemployment-rates-oecd-update-october-2021.htm
}), crude oil prices west texas intermediate (WTI, \url{https://fred.stlouisfed.org/series/MCOILWTICO
}) and the quarterly GDP (\url{https://data.oecd.org/gdp/quarterly-gdp.htm#indicator-chart
}) are employed from 2000 to 2020. The codes and cleaning data are available on GitHub: \url{https://github.com/kannyjyk/TVP-GVAR-ML}. We select the top three advanced economies as our targets: the United States, Euro Area, and Japan.

\subsection{Basic model}
\indent\indent As was previously mentioned, our economic inputs come at different frequencies. In light of this, the imputation of economic variables at the same time intervals in advance is carried out in accordance with \citet{ankargren2020flexible}. The presented approaches concentrate on $p$ economic variables, which are supposed to be endogenous variables for each country. Consider exogenous variables from the common economic activity (behaviour) and the remaining countries, and there are $K$ countries at moment $t$.  We also make the simplifying assumption that all exogenous variables have first-order lag. For each country $k=1,...,K$, there is a time-varying parameter GVAR (TVP-GVAR) model given by
\begin{equation}\label{equ:3}
\boldsymbol{x}_{k, t}^{(E)}=\boldsymbol{a}_{k}^{(E)}+\mathbf{\Phi}_{1}^{(E, E)} \boldsymbol{x}_{k, t-1}^{(E)}+\sum_{j=0}^{1} \mathbf{\Gamma}_{k j}^{(E, E)} \boldsymbol{x}_{k, t - j}^{*(E, E)}+ \sum_{j=0}^{1} \mathbf{\Gamma}_{k j}^{(E, B)} \boldsymbol{x}_{k, t - j}^{*(E, B)}+\boldsymbol{u}_{k, t}^{(E)},
\end{equation}
where the superscripts $E$ and $B$ denote an economy sovereign/country and a common economic activity, respectively. $*$ represents an exogenous form of the variables. $\boldsymbol{a}_{k}^{(E)}$ is a $p$-dimensional constant vector, error term $\boldsymbol{u}_{k, t}^{(E)}$ is uncorrelated each other, and cross-sectional weak independent with expectation $\boldsymbol{0}$ and $\bold{\Sigma}_{\boldsymbol{u}}=Var(\boldsymbol{u}_{k, t})$. 


For each country $k$, there are $$\boldsymbol{x}_{k,t}^{(E)} = \left[\begin{array}{c}
x_{k 1, t}^{(E)} \\
x_{k 2, t}^{(E)} \\
\vdots \\
x_{k p, t}^{(E)}
\end{array}\right] \in \mathbb{R}^{p \times 1}, \quad
\mathbf{X}_t^{(E)} = \left[\begin{array}{c}
{\boldsymbol{x}_{1,t}^{(E)}}^\top \\
{\boldsymbol{x}_{2,t}^{(E)}}^\top \\
\vdots \\
{\boldsymbol{x}_{K,t}^{(E)}}^\top
\end{array}\right] \in \mathbb{R}^{K \times p}, \quad
k=1,2, \cdots, K.$$
Also, as country-wise interdependent, there are $p+l$ exogenous variables for country $k$ out of all $K$ countries: 
$$\boldsymbol{x}_{k,t}^{*(E)}=\left[\begin{array}{c}
\boldsymbol{x}_{k, t}^{* (E, E)} \\
\boldsymbol{x}_{k, t}^{* (E, B)}
\end{array}\right] \in \mathbb{R}^{(p+l) \times 1}.$$
For $p$ variables from one of the other countries, there are $$\boldsymbol{x}_{k,t}^{* (E, E)} = \left[\begin{array}{c}
x_{k 1, t}^{*(E,E)} \\
x_{k 2, t}^{*(E,E)} \\
\vdots \\
x_{k p, t}^{*(E,E)}
\end{array}\right] \in \mathbb{R}^{p \times 1}, \quad
\mathbf{X}_t^{* (E, E)} = \left[\begin{array}{c}
{\boldsymbol{x}_{1,t}^{* (E, E)}}^\top \\
{\boldsymbol{x}_{2,t}^{* (E, E)}}^\top \\
\vdots \\
{\boldsymbol{x}_{K,t}^{* (E, E)}}^\top
\end{array}\right] \in \mathbb{R}^{K \times p},$$
where $$\mathbf{X}_t^{* (E, E)} = \mathbf{W}_t^{(E)} \mathbf{X}_t^{(E)}, \quad  w_{ii,t}^{(E)} = 0, \quad \sum_{i=1}^K w_{ik,t}^{(E)} = 1.$$
$\mathbf{W}_t^{(E)}$ is a $K \times K$ weight matrix at moment $t$, an element $w_{ik,t}^{(E)}$ is defined as transform weight for a country $k$ at $k$-th column in a matrix.   

Additionally, supposed to all $K$ country economies are affected by $l$ economic behaviours, i.e., there are $l$ common external shocks for all countries. We then have   
$\boldsymbol{x}_{k,t}^{*(E,B)} \equiv \boldsymbol{x}_{t}^{(B)}, k = 1, 2, \cdots, K$, $\boldsymbol{x}_{t}^{(B)}=\left[x_{1, t}^{(B)}, x_{2, t}^{(B)},...,x_{l, t}^{(B)}\right]^\top$ represent $l$ common economic activities. 

For each $x_{l, t}^{(B)}$, it has a similar form to $\boldsymbol{x}_{k, t}^{(E)}$, which holds
\begin{equation}\label{equ:4}
x_{l, t}^{(B)}=a_{l}^{(B)}+\Phi_l^{(B, B)} x_{l, t-1}^{(B)}+\sum_{j=0}^{1} \mathbf{\Gamma}_{lj}^{(B, E)} \boldsymbol{x}_{l, t-j}^{*(B, E)}+u_{l, t}^{(B)},
\end{equation}
where $\{\boldsymbol{u}_{t}^{(B)}\}$ is again uncorrelated each other and weak stationary with expectation 
$\boldsymbol{0}$. $\boldsymbol{x}_{t}^{*(B, E)} = \left[\boldsymbol{x}_{1, t}^{*(B, E)}, \boldsymbol{x}_{2, t}^{*(B, E)},...,\boldsymbol{x}_{l, t}^{*(B, E)}\right]_{p \times l}$. The $l$ exogenous variables for common economic activities from countries are defined as $$\boldsymbol{x}_{t}^{*(B, E)} = {\mathbf{X}_t^{(E)}}^\top \mathbf{W}_t^{(B)} \in \mathbb{R}^{p \times l},$$
where $$\mathbf{W}_t^{(B)} = \left[\begin{array}{cccc}
w_{1 1, t}^{(B)} & w_{1 2, t}^{(B)}&\cdots&w_{1 l, t}^{(B)}  \\
\vdots &\cdots&\cdots&\vdots \\
w_{K 1, t}^{(B)} & w_{K 2, t}^{(B)} &\cdots&w_{K l, t}^{(B)}  
\end{array}\right] \in \mathbb{R}^{K \times l}$$ is a weight matrix for all $K$ countries to common economic activities. A $k$ country may be involved $l$ common economic activities, the weights could be satisfied by $\sum_{k=1}^K w_{km,t} = 1,\  m = 1, 2,\cdots,l$.

In practice, based on two weight matrices $\mathbf{W}_t^{(E)}$ and $\mathbf{W}_t^{(B)}$, we find out the solutions for $\boldsymbol{x}_{k, t - j}^{*(E, E)}, \boldsymbol{x}_{k, t - j}^{*(E, B)}$ in Equation (\ref{equ:3}) and $\boldsymbol{x}_{t-j}^{*(B, E)}$ in Equation (\ref{equ:4}). Then, we estimate the coefficients for $\mathbf{\Phi}_{1}^{(E, E)}$, $\boldsymbol{\Phi}^{(B, B)}$, $\boldsymbol{\Gamma}_{k j}^{(E, E)}$, $\mathbf{\Gamma}_{k j}^{(E, B)}$, $\mathbf{\Gamma}_{l j}^{(B, E)}$ by the least squares method, where $j = 0,1$. To facilitate the calculation, we firstly conduct the variable substitution, which defines as:
\begin{align*}
\boldsymbol{z}_{k, t}^{(E)} & = \left[\begin{array}{c}
\boldsymbol{x}_{k, t}^{(E)} \\
\boldsymbol{x}_{k, t}^{*(E)}
\end{array}\right] =  \left[\begin{array}{c}
\boldsymbol{x}_{k, t}^{(E)} \\
\boldsymbol{x}_{k, t}^{*(E, E)} \\
\boldsymbol{x}_{t}^{(B)} 
\end{array}\right]_{(2p+l) \times 1} \\
& = \begin{array}{c}
1 \quad \cdots \quad\quad k \quad\quad \cdots  \quad\quad K \quad\quad\quad \\
\left[\begin{array}{cccccc}
\boldsymbol{0} & \cdots & \mathbf{I}_p & \cdots & \boldsymbol{0}   & \boldsymbol{0} \\
w_{1k,t}^{(E)} \mathbf{I}_p & \cdots & w_{kk,t}^{(E)} \mathbf{I}_p & \cdots& w_{Kk,t}^{(E)} \mathbf{I}_p & \boldsymbol{0} \\
\boldsymbol{0} & \cdots & \boldsymbol{0} & \cdots & \boldsymbol{0} & \mathbf{I}_l
\end{array}\right]
\end{array}_{(2p+l) \times (Kp+l)} \cdot \left[\begin{array}{c}
\boldsymbol{x}_{1, t}^{(E)} \\
\vdots \\
\boldsymbol{x}_{K, t}^{(E)} \\
\boldsymbol{x}_{t}^{(B)} 
\end{array}\right]_{(Kp+l) \times 1} \\
& =: \boldsymbol{L}_{k,t}^{(E)} \boldsymbol{x}_t.
\end{align*}
Thus, for each country $k$, the basic model (Equation (\ref{equ:3})) can be transformed as:
\begin{equation}\label{equ:5}
\begin{aligned}
& \boldsymbol{A}_{k 0}^{(E)} \boldsymbol{z}_{k,t}^{(E)}=\boldsymbol{a}_{k}^{(E)}+\boldsymbol{A}_{k 1}^{(E)} \boldsymbol{z}_{k,t-1}^{(E)}+\boldsymbol{u}_{k, t}^{(E)} \\
\Leftrightarrow \quad & \boldsymbol{A}_{k 0}^{(E)} \boldsymbol{L}_{k,t}^{(E)} \boldsymbol{x}_{t}=\boldsymbol{a}_{k}^{(E)}+\boldsymbol{A}_{k 1}^{(E)} \boldsymbol{L}_{k,t-1}^{(E)} \boldsymbol{x}_{t-1}+\boldsymbol{u}_{k, t}^{(E)}.
\end{aligned}
\end{equation}  
Then, Equation (\ref{equ:3}) has the form as:
\begin{equation*}
\boldsymbol{x}_{k, t}^{(E)}=\boldsymbol{a}_{k}^{(E)}+\mathbf{\Phi}_{1}^{(E, E)} \boldsymbol{x}_{k, t-1}^{(E)}+\sum_{j=0}^{1} \left[ \boldsymbol{\Gamma}_{k j}^{(E, E)}, \boldsymbol{\Gamma}_{k j}^{(E, B)} \right] \left[\begin{array}{c}
\boldsymbol{x}_{k, t - j}^{*(E, E)} \\
\boldsymbol{x}_{k, t - j}^{*(E, B)} 
\end{array}\right] + \boldsymbol{u}_{k, t}^{(E)},
\end{equation*}
and Equation (\ref{equ:5}) is equivalent to the following forms:
$$\boldsymbol{A}_{k0}^{(E)} = \left[ \mathbf{I}_p, -\mathbf{\Gamma}_{k 0}^{(E, E)}, -\mathbf{\Gamma}_{k 0}^{(E, B)}\right] \in \mathbb{R}^{p \times (2p+l)},$$
$$\boldsymbol{A}_{k1}^{(E)} = \left[ \mathbf{\Phi}_{1}^{(E, E)}, \mathbf{\Gamma}_{k 1}^{(E, E)}, \mathbf{\Gamma}_{k 1}^{(E, B)}\right] \in \mathbb{R}^{p \times (2p+l)}.$$ 
We have $K$ countries, and thus Equation (\ref{equ:5}) has a stacked form as:
$$\boldsymbol{G}_{0,t}^{(E)} \boldsymbol{x}_{t}=\boldsymbol{a}^{(E)}+\boldsymbol{G}_{1,t-1}^{(E)} \boldsymbol{x}_{t-1}+\boldsymbol{u}_{t}^{(E)},$$
where
\begin{equation*}
\begin{aligned}
\boldsymbol{G}_{0,t}^{(E)}=\left[\begin{array}{c}
\boldsymbol{A}_{10}^{(E)} \boldsymbol{L}_{1,t}^{(E)} \\
\boldsymbol{A}_{20}^{(E)} \boldsymbol{L}_{2,t}^{(E)} \\
\vdots \\
\boldsymbol{A}_{K 0}^{(E)} \boldsymbol{L}_{K,t}^{(E)}
\end{array}\right]_{Kp \times (Kp + l)}, 
\boldsymbol{G}_{1,t}^{(E)}=\left[\begin{array}{c}
\boldsymbol{A}_{1 1}^{(E)} \boldsymbol{L}_{1,t}^{(E)} \\
\boldsymbol{A}_{2 1}^{(E)} \boldsymbol{L}_{2,t}^{(E)} \\
\vdots \\
\boldsymbol{A}_{K 1}^{(E)} \boldsymbol{L}_{K,t}^{(E)}
\end{array}\right]_{Kp \times (Kp + l)}, 
\\
\boldsymbol{a}^{(E)}=\left[\begin{array}{c}
\boldsymbol{a}_{1}^{(E)} \\
\boldsymbol{a}_{2}^{(E)} \\
\vdots \\
\boldsymbol{a}_{K}^{(E)}
\end{array}\right]_{Kp \times 1}, 
\boldsymbol{u}_{t}^{(E)}=\left[\begin{array}{c}
\boldsymbol{u}_{1, t}^{(E)} \\
\boldsymbol{u}_{2, t}^{(E)} \\
\vdots \\
\boldsymbol{u}_{K, t}^{(E)}
\end{array}\right]_{Kp \times 1}.
\end{aligned}
\end{equation*} 

Next, we consider $l$ common economic activities. Equation (\ref{equ:4}) can have a similar form to the above process, denoting:
\begin{equation}\label{equ:6}
\begin{aligned}
& \boldsymbol{A}_{0}^{(B)} \boldsymbol{z}_{t}^{(B)}=\boldsymbol{a}^{(B)}+\boldsymbol{A}_{1}^{(B)} \boldsymbol{z}_{t-1}^{(B)}+\boldsymbol{u}_{t}^{(B)} \\
\Leftrightarrow \quad & \boldsymbol{A}_{0}^{(B)} \boldsymbol{L}_{t}^{(B)} \boldsymbol{x}_{t}=\boldsymbol{a}^{(B)}+\boldsymbol{A}_{1}^{(B)} \boldsymbol{L}_{t-1}^{(B)} \boldsymbol{x}_{t-1}+\boldsymbol{u}_{t}^{(B)},
\end{aligned}
\end{equation} 
where $\boldsymbol{z}_{t}^{(B)}=\left[\boldsymbol{z}_{1, t}^{(B)}, \boldsymbol{z}_{2, t}^{(B)}, \cdots, \boldsymbol{z}_{l, t}^{(B)}\right]^\top,\ \boldsymbol{L}_{t}^{(B)}=\left[\boldsymbol{L}_{1, t}^{(B)}, \boldsymbol{L}_{2, t}^{(B)}, \cdots, \boldsymbol{L}_{l, t}^{(B)}\right]^\top$, 
\begin{align*}
\boldsymbol{z}_{m, t}^{(B)} & = \left[\begin{array}{c}
\boldsymbol{x}_{m, t}^{(B)} \\
\boldsymbol{x}_{m, t}^{*(B)}
\end{array}\right]_{(p+1) \times 1} \\
& = \begin{array}{c}
\quad\quad\quad\quad\quad\quad\quad\quad\quad\quad 1 \quad \cdots \ m \ \cdots \quad l \\
\left[\begin{array}{cccccccc}
\boldsymbol{0} & \cdots & \boldsymbol{0}  & 0 & \cdots & 1 & \cdots & 0 \\
w_{11,t}^{(B)} \mathbf{I}_p & \cdots& w_{K1,t}^{(B)} \mathbf{I}_p & \boldsymbol{0} & \cdots & \boldsymbol{0} & \cdots & \boldsymbol{0}
\end{array}\right]
\end{array}_{(p+1) \times (Kp+l)} \cdot \left[\begin{array}{c}
\boldsymbol{x}_{1, t}^{(E)} \\
\vdots \\
\boldsymbol{x}_{K, t}^{(E)} \\
\boldsymbol{x}_{t}^{(B)} 
\end{array}\right]_{(Kp+l) \times 1} \\
& =: \boldsymbol{L}_{m, t}^{(B)} \boldsymbol{x}_t.
\end{align*}
Combined Equations (\ref{equ:4}) and (\ref{equ:6}), we have 
$$\boldsymbol{A}_{m0}^{(B)} = \left[ 1, -\mathbf{\Gamma}_{m 0}^{(B, E)}\right] \in \mathbb{R}^{1 \times (p+1)},$$
$$\boldsymbol{A}_{m1}^{(B)} = \left[ \Phi_{l}^{(B, B)}, \mathbf{\Gamma}_{m 1}^{(B, E)}\right] \in \mathbb{R}^{1 \times (p+1)},$$
where $m=1,2,\cdots,l$ and $\boldsymbol{A}_{j}^{(B)}=\left[\boldsymbol{A}_{1, j}^{(B)}, \boldsymbol{A}_{2, j}^{(B)}, \cdots, \boldsymbol{A}_{l, j}^{(B)}\right]^\top$, $j = 0,1$. Equation (\ref{equ:6}) has a form as:
$$\boldsymbol{G}_{0,t}^{(B)} \boldsymbol{x}_{t}=\boldsymbol{a}^{(B)}+\boldsymbol{G}_{1,t-1}^{(B)} \boldsymbol{x}_{t-1}+\boldsymbol{u}_{t}^{(B)},$$
where
$$\boldsymbol{G}_{0,t}^{(B)}=\left[\begin{array}{c}
\boldsymbol{A}_{10}^{(B)} \boldsymbol{L}_{1, t}^{(B)} \\
\vdots \\
\boldsymbol{A}_{l0}^{(B)} \boldsymbol{L}_{l, t}^{(B)}
\end{array}\right]_{l \times 1} ,
\boldsymbol{G}_{1,t}^{(B)}=\left[\begin{array}{c}
\boldsymbol{A}_{11}^{(B)} \boldsymbol{L}_{1, t}^{(B)} \\
\vdots \\
\boldsymbol{A}_{l1}^{(B)} \boldsymbol{L}_{l, t}^{(B)}
\end{array}\right]_{l \times 1} .$$
We consider Equations (\ref{equ:3}) and (\ref{equ:4}) together, and the coefficient matrices can be obtained as:
\begin{equation*}
\begin{aligned}
\boldsymbol{G}_{0,t}=\left[\begin{array}{c}
\boldsymbol{G}_{0,t}^{(E)} \\
\boldsymbol{G}_{0,t}^{(B)}
\end{array}\right]_{(Kp + l) \times (Kp + l)}, \boldsymbol{G}_{1,t}=\left[\begin{array}{c}
\boldsymbol{G}_{1,t}^{(E)} \\
\boldsymbol{G}_{1,t}^{(B)}
\end{array}\right]_{(Kp + l) \times (Kp + l)}, 
\\
\boldsymbol{a}=\left[\begin{array}{c}
\boldsymbol{a}_{1}^{(E)} \\
\boldsymbol{a}_{2}^{(E)} \\
\vdots \\
\boldsymbol{a}_{K}^{(E)} \\
\boldsymbol{a}^{(B)}
\end{array}\right]_{(Kp + l) \times 1}, \boldsymbol{u}_{t}==\left[\begin{array}{c}
\boldsymbol{u}_{1, t}^{(E)} \\
\boldsymbol{u}_{2, t}^{(E)} \\
\vdots \\
\boldsymbol{u}_{K, t}^{(E)} \\
\boldsymbol{u}_{t}^{(B)}
\end{array}\right]_{(Kp + l) \times 1}.
\end{aligned}
\end{equation*}

Now, we combine Equation (\ref{equ:3}) and Equation (\ref{equ:4}) to obtain
\begin{equation}\label{equ:7}
\boldsymbol{x}_{t}=\boldsymbol{b}_{t}+\boldsymbol{F}_{1,t-1} \boldsymbol{x}_{t-1}+\boldsymbol{\varepsilon}_{t},
\end{equation}
where
$$
\boldsymbol{b}_{t}=\boldsymbol{G}_{0, t}^{-1} \boldsymbol{a},\quad \boldsymbol{F}_{1, t}=\boldsymbol{G}_{0, t}^{-1} \boldsymbol{G}_{1, t},\quad \boldsymbol{\varepsilon}_{t}=\boldsymbol{G}_{0, t}^{-1} \boldsymbol{u}_{t}.
$$

Then, we again employ the moving average process for Equation (\ref{equ:7}),
$$\boldsymbol{x}_{t}=\boldsymbol{d}_{t}+\sum_{s=0}^{\infty} \boldsymbol{B}_{s} \boldsymbol{\varepsilon}_{t-s},$$
where $\boldsymbol{d}_{t}$ denotes the determinant part of $\boldsymbol{x}_{t}$, and $\boldsymbol{B}_{s}$ is constructed
$$\boldsymbol{B}_{s} =\boldsymbol{F}_{1,t} \boldsymbol{B}_{s-1}, s=1,2, \ldots,$$
in which $\boldsymbol{B}_{0} =\mathbf{I}_{Kp+l}$, for $s<0$, there exists $\boldsymbol{B}_{s}=\mathbf{0}$.


\subsection{Time-varying parameters (TVPs) estimations}
\indent\indent For each variable in every country $k$, there is a compact version of Equation (\ref{equ:7}), 
\begin{equation} \label{equ:full}
x_{it}=b_{it}+\textit{F}_{1i,t-1} x_{i,t-1}+\varepsilon_{it}, \quad i=k, \ldots, Kp+l.
\end{equation}

For convenience, we define $N = Kp+l$. Allowing for TVPs is straightforward by drawing from the vast literature on state-space models. The regression coefficients are stacked in a $2 \times 1$ -vector $\boldsymbol{\theta}_{t}=(b_{it}, f_{1it})^{\prime}.$ We assume independent random walk state equations
	$$
	\boldsymbol{\theta}_{it}=\boldsymbol{\theta}_{it-1}+\boldsymbol{\eta}_{it}, \quad \boldsymbol{\eta}_{it} \sim \mathcal{N}\left(\mathbf{0}, \boldsymbol{\Omega}_{i}\right).
	$$
	Here, $\boldsymbol{\eta}_{i t}$ is a zero-mean Gaussian error term and diagonal covariance matrix $\boldsymbol{\Omega}=\operatorname{diag}(\omega_{ i1}, \omega_{ i2})$ of size $2 \times 2$. The state innovation variances in $\boldsymbol{\Omega}$ govern the degree of time variation in the regression coefficients, where we develop the following algorithm based on \citet{hauzenberger2021bayesian}.  

\begin{breakablealgorithm}
\caption{The estimations of parameters in TVP-GVAR}
\begin{algorithmic}
1. Initialize the $\boldsymbol{\theta}_{i0}=0^{2\times 1},\boldsymbol{\Omega}_i= \text{diag}(1, 1),\Sigma=\text{diag}(\sigma_1^2, \cdots, \sigma_N^2)=\text{diag}(0.1, \cdots, 0.1), i = 1,\cdots,N$. \\
\indent\indent 2. $\tilde{\boldsymbol{\theta}}_{t}$, to define the prior distribution on the time-varying regression coefficients, we consider the state-space model in its non-centred parametrization. Let $\sqrt{\boldsymbol{\Omega}}_i=\operatorname{diag}(\sqrt{\omega_{i1}}, \sqrt{\omega_{i2}})$, and then we split the coefficients into a constant and time-varying part: 
		$$\boldsymbol{\theta}_{i t}=\boldsymbol{\theta}_{i0}+\sqrt{\boldsymbol{\Omega}}_i \tilde{\boldsymbol{\theta}}_{it}.$$
		
Using this transformation, $\tilde{\boldsymbol{\theta}}_{t}$ follows a random walk with standard normal shocks and is generated by the Kalman filter,
\begin{equation} \label{equ:8}
y_{i t}-\left[1, y_{i, t-1}\right] \theta_{i 0}=\left[1, y_{i, t-1}\right] \sqrt{\boldsymbol{\Omega}}_i \tilde{\theta}_{i t}+\epsilon_{i t}, \quad \epsilon_{i t} \sim \mathcal{N}(0, \sigma_i^2).
\end{equation}

We set $\tilde{\boldsymbol{\theta}}_{i0} \sim \mathrm{N}(\mathbf{m}_{i0}, \mathbf{P}_{i0},\mathbf{R}=0.1^{2\times 1},\mathbf{m}_{i0}=0^{2\times 1},\mathbf{P}_{i0}=10^{-15}\textbf{I}_{2},\textbf{H}=[1,y_{i,t-1}],y_{it}^{\star}=y_{it}-[1,y_{i,t-1}]\boldsymbol{\theta}_{i0}$ initially. And the iteration formulas are as follows:\\
		$$
		\begin{aligned}
		\mathbf{P}_{it}^{-} &=\mathbf{P}_{it-1}+\mathbf{Q}\\
		\mathbf{v}_{it}&=y_{it}^{\star}-\mathbf{H} \mathbf{m}_{it-1}, \\
		\mathbf{S}_{it} &=\mathbf{H} \mathbf{P}_{it}^{-} \mathbf{H}^{\top}+\mathbf{R}, \\
		\mathbf{K}_{it} &=\mathbf{P}_{it}^{-} \mathbf{H}^{\top} \mathbf{S}_{t}^{-1},\\
		\mathbf{m}_{it} &=\mathbf{m}_{it-1}+\mathbf{K}_{it} \mathbf{v}_{it}, \\
		\mathbf{P}_{it} &=\mathbf{P}_{it}^{-}-\mathbf{K}_{it} \mathbf{S}_{it} \mathbf{K}_{it}^{\top}.
		\end{aligned}
		$$
	
	After we have expectations and variances, we can sample $\tilde{\boldsymbol{\theta}}_{t}$ from $\mathrm{N}(\mathbf{m}_{it}, \mathbf{P}_{i0})$.\\
\indent\indent 3. $\boldsymbol{\theta}_{i0},\sqrt{\boldsymbol{\Omega}}_i$, change the Formula (\ref{equ:8}):
	$$
	y_{it}=[1,y_{i,t-1},[1,y_{i,t-1}]\tilde{\boldsymbol{\theta}}_{it}]\left[\boldsymbol{\theta}_{i0},\left[\sqrt{\omega_{i1}}, \sqrt{\omega_{i2}}\right]\right]'+\epsilon_{it}, \quad \epsilon_{it} \sim \mathcal{N}(0, \sigma_i^{2}).
	$$
	
Set $\textbf{y}_{i,t-1}^{\star}=[1,y_{i,t-1},[1,y_{i,t-1}']\tilde{\boldsymbol{\theta}}_{it}]/\sigma_i,\quad\textbf{y}_{i}^{\star}=\left[\textbf{y}_{i,1}^{\star}, \cdots, \textbf{y}_{i,T-1}^{\star}\right],\quad\boldsymbol{\theta}_{i}^{\star}=[\boldsymbol{\theta}_{i0},\left[\sqrt{\omega_{i1}}, \sqrt{\omega_{i2}}\right]]'$. Therefore, $\boldsymbol{\theta}_{i}^{\star}$ can be sampled according to the practices in the paper of \citet{fruhwirth2010stochastic}, in which $\boldsymbol{\theta}_{i}^{\star}$ join from the normal posterior $\mathrm{N}(\textbf{a}_{i},\textbf{A}_{i})$
	$$
	\begin{aligned}
		\mathbf{A}_{i}&=((\textbf{y}^{\star}_{i})^{\prime} \textbf{y}^{\star}_{i}+(\mathbf{A}_{i0})^{-1})^{-1} \\
		\mathbf{a}_{i}&=\mathbf{A}_{i}\left((\textbf{y}^{\star}_{i})^{\prime} \mathbf{y_{i}}+(\mathbf{A}_{i0})^{-1} \mathbf{a}_{i0}\right),
			\end{aligned}
				$$
	where $\mathbf{a}_{i0}=0^{2\times 1},\mathbf{A}_{i0}=\text{diag}\{1/\text{diag}((\bar{y}'\bar{y})^{-1})\}$.\\
\indent\indent 4. $\Sigma=\text{diag}\{\sigma_1,\dots ,\sigma_N\}$, it is mentioned in \citet{fruhwirth2010stochastic}'s article that $\sigma_{i}$ is subject to gamma distribution:
	
		$$
		\begin{aligned}
		&\sigma_{i}\sim\mathrm{G}(c_{iT},C_{iT}) \\
		&c_{iT}=c_{i0}+T / 2, \\
		&C_{iT}=C_{i0}+\frac{1}{2}(\mathbf{y_i}^{\prime}-\textbf{y}_{i}^{\star}\boldsymbol{\theta}_{i}^{\star})^2.
	\end{aligned}
	$$
	Thus $\sigma_{i}$ can be sampled from the above distribution.\\
\indent\indent 5. Repeat the above process from 1 to 3, and take the sampling result of the last iteration as the estimation of the parameters to be estimated.\\
\indent\indent 6. Finally, an estimate for $\boldsymbol{\theta}_{it}$ can be calculated by substituting the $\boldsymbol{\theta}_{i0}$ and $\tilde{\boldsymbol{\theta}}_{it}$ obtained from each estimate into the Formula $\boldsymbol{\theta}_{i t}=\boldsymbol{\theta}_{i0}+\sqrt{\boldsymbol{\Omega}}_i \tilde{\boldsymbol{\theta}}_{it} $.
\end{algorithmic}
\end{breakablealgorithm}

\subsection{IRF}
\indent\indent To identify the response of a (common) shock, the existing IRF has been developed to provide estimates. We aim to offer reasonable estimates inferred from the observed data and interpret targeting responses to an emerging event. Empirically, IRF may have the general form, called general IRF (GIRF), in which multiple variables at future moment $t+n$ are affected by a shock at current time $t$, 
$$
\mathbf{GIRF} \left(\boldsymbol{x}_{t} ; u_{k \ell t}, n\right)=\mathbf{E}\left(\boldsymbol{x}_{t+n} \mid u_{k \ell t}=\sqrt{\sigma_{k k, \ell \ell}}, \mathcal{I}_{t-1}\right)-\mathbf{E}\left(\boldsymbol{x}_{t+n} \mid \mathcal{I}_{t-1}\right),
$$
where $\mathcal{I}_{t-1}$ is the information set at the moment $t-1$, $\sigma_{k k, \ell \ell}$ are diagonal elements in a covariance matrix $\mathbf{\Sigma}_{u}$, $\mathbf{\Sigma}_{u}= \mathbf{E}(\boldsymbol{u}_t \boldsymbol{u}_t^\top)$, corresponding to $\ell$-th equations for a country $k$, and  $n$ is the length of the observational period. Here, we take a further look at the orthogonal IRF (OIRF) for the numerous economic variables' impacts on an economic variable.

%

\subsubsection{\textbf{TVP-OIRF}}:
Consider Cholesky decomposition for a covariance matrix $\boldsymbol{\Sigma}_{u}$, we have
$$
\mathbf{P P}^{\top}=\boldsymbol{\Sigma}_{u},
$$
where $\mathbf{P}$ is a $\left(Kp+l\right) \times \left(Kp+l\right)$ lower triangular matrix, then Equation (\ref{equ:7}) may have the form as
$$
\begin{aligned}
\boldsymbol{x}_{t}&=\boldsymbol{d}_t+\sum_{s=0}^{\infty} \mathbf{B}_{s} \left( \boldsymbol{G}_{0,t}^{-1} \boldsymbol{u}_{t-s}\right)\\
\boldsymbol{x}_{t}&=\boldsymbol{d}_t+\sum_{s=0}^{\infty} \left(\mathbf{B}_{s} \boldsymbol{G}_{0,t}^{-1} \mathbf{P} \right)\left( \mathbf{P}^{-1} \boldsymbol{u}_{t-s}\right)\\
\boldsymbol{x}_{t}&=\boldsymbol{d}_t+\sum_{s=0}^{\infty} \left(\mathbf{B}_{s} \boldsymbol{G}_{0,t}^{-1} \mathbf{P} \right)\boldsymbol{\xi}_{t-s},
\end{aligned}
$$
where $\boldsymbol{\xi}_{t}=\mathbf{P}^{-1} \boldsymbol{u}_{t}$ is orthogonal, i.e., $\mathbf{E}\left(\boldsymbol{\xi}_{t}\boldsymbol{\xi}_{t}^{\top}\right)=\mathbf{I}_{\left(Kp+l\right)}$. Thus, for $j$-th variable (a variable for a country $k$ in a GVAR model) given a unit impulse $\delta_j$ at the moment $t$, it could have orthogonal IRF at the moment $t+n$.  
$$
\mathbf{OImp}_j(n)=\mathbf{B}_{n} \boldsymbol{G}_{0,t}^{-1} \mathbf{P} \boldsymbol{e}_j, n = 0,1,2,\cdots,
$$
where $\mathbf{e}_j$ is a $\left(Kp+l\right) \times 1$ column vector, with $j$-th element as 1 and others are 0. Note that we will abbreviate it as OIRF below.

The OIRF may involve shocks for $p$ variables, which sum up all corresponding OIRF of each variable. This follows by the orthogonal property of impulse,  
$$
\mathbf{OImp}_{j_1, \cdots, j_k}(n)=\mathbf{B}_{n} \boldsymbol{G}_{0,t}^{-1} \mathbf{P} \left(\mathbf{e}_{j_1} + \cdots + \mathbf{e}_{j_k} \right), n = 0,1,2,\cdots.
$$

\subsubsection{\textbf{Asymptotic distributions}:}

In order to calculate the error bands of the orthogonal impulse response, we introduce the asymptotic distribution of it proposed by \citet{lutkepohl2020constructing} and \citet{lutkepohl2005new}.
\begin{equation} \label{equ:asym}
	\sqrt{T}\ \text{vec}(\widehat{\mathbf{OImp}}(n) - \mathbf{OImp}(n)) \stackrel{d}{\rightarrow} \mathcal{N}(0,\bm{C}_n\bm{\Sigma}_{\widehat{\bm{\alpha}}} \bm{C}^\top_n + \bm{\bar{C}}_n\bm{\Sigma}_{\widehat{\bm{\sigma}}}\bar{\bm{C}}^\top_n),
\end{equation}
where $\mathbf{OImp}(n) = [\mathbf{OImp}_1(n)^\top,\ldots, \mathbf{OImp}_{Kp+l}(n)^\top]^\top$ and the total period is $T$.
$$
\begin{aligned}
	& \bm{\alpha} := \text{vec}(\bm{F}_{1}),\quad \bm{\sigma} := \text{vech}(\bm{\Sigma}_{\bm{\varepsilon}} )
	\\
	& \bm{C}_0 := 0,\quad \bm{C}_n := (\bm{P}_\varepsilon^\top\otimes \mathbf{I}_{Kp+l})\bm{G}_n
	\\
	& \bar{\bm{C}}_n :=(\mathbf{I}_{Kp+l} \otimes \bm{B}_n)\bm{H}
	\\
	& \bm{H}:=\partial\text{vec}(\bm{P}_\varepsilon) / \partial\bm{\sigma}^\top = \bm{L}^\top_{Kp+l}\{\bm{L}_{Kp+l}(\mathbf{I}_{(Kp+l)^2}+\bm{K}_{Kp+l,Kp+l}) (\bm{P}_\varepsilon\otimes \mathbf{I}_{Kp+l})\bm{L}^\top_{Kp+l}\}^{-1},
\end{aligned}
$$
where $\bm{P}_\varepsilon$ is the Cholesky decomposition of $\bm{\Sigma}_{\bm{\varepsilon}}$ and $\bm{G}_n := \partial\text{vec}(\bm{B}_n)/\partial\bm{\alpha}^\top = \sum^{n-1}_{m=0}(\bm{F}_1^\top)^{n-1-m} \otimes \bm{B}_m$. The communication matrix $\bm{K}_{mn}$ is defined such that, for any $(m\times n)$ matrix $\bm{Q}$, $\bm{K}_{mn}\text{vec}(\bm{Q}) = \text{vec}(\bm{Q}^\top)$. Furthermore, $\bm{L}_m$ is the $(m(m+1)/2\times m^2)$ elimination matrix defined such that, for any $(m\times m)$ matrix $\bm{S}$, $\text{vech}(\bm{S}) = \bm{L}_m\text{vec}(\bm{S})$. And vec($\cdot$) is the column stacking operator, and vech($\cdot$) is the column stacking operator for symmetric matrices (stacks the elements on and below the main diagonal only). $\bm{A}\otimes\bm{B}$ represents the Kronecker product of matrices $\bm{A}$ and $\bm{B}$.

Then the error bands can be formed as the result of (\ref{equ:asym}):
$$
P\left(\widehat{\text{OImp}}_{ij}(n) - z_{1-\alpha/2}\frac{1}{\sqrt{T}}\hat{\sigma}(\widehat{\text{OImp}}_{ij}) \leq \text{OImp}_{ij} \leq \widehat{\text{OImp}}_{ij}(n) + z_{1-\alpha/2}\frac{1}{\sqrt{T}}\hat{\sigma}(\widehat{\text{OImp}}_{ij})\right)
=1-\alpha,
$$
where $z_{1-\alpha/2}$ denotes the $1-\alpha/2$-quantile of the standard normal distribution and $\widehat{\text{OImp}}_{ij}$ is the $i$-th element of $\mathbf{OImp}_j(n)$ and $\hat{\sigma}(\widehat{\text{OImp}}_{ij})$ is the square root of element $((Kp+l)(j-1)+i,(Kp+l)(j-1)+i)$ of the estimated $(\bm{C}_n\bm{\Sigma}_{\widehat{\bm{\alpha}}} \bm{C}^\top_n + \bm{\bar{C}}_n\bm{\Sigma}_{\widehat{\bm{\sigma}}}\bar{\bm{C}}^\top_n)$.

\subsection{Forecasting}
\indent\indent There has been increasing use of various machine learning models in forecasting. To address the interpretability of estimates, we obtain empirical measures of estimates from the TVP process as training sets and explore their time-varying values until the end time using machine learning models. Once the time-varying estimates have been produced, we allow those values back to the TVP process for the final forecasting economic variables. We consider the time-varying weights of the training data on the last day and fix them to estimate the time-varying values of the testing data, called constant prediction. The estimation also considers the Vector autoregressive regression (VAR) \citep{sims1980macroeconomics,johansen1995likelihood} process. The other existing models including TREE, Random Forest (RF) \citep{breiman2001random}, Least Absolute Shrinkage and Selection Operator (LASSO) \citep{tibshirani1996regression,tibshirani2011regression}, Long Short Term Memory (LSTM) \citep{10.1162/neco.1997.9.8.1735},  Gated Recurrent Unit (GRU) \citep{chung2014empirical}, and LSTM combined with GRU (LSTM\_GRU)  are used to implement time-varying estimates. Then, we use these estimates to project the following trajectories of economic variables. The projection spans from January to June of 2021. Finally, the prediction performance is calculated using Mean Squared Error (MSE):
$$MSE= \sum^t_{i=1}\frac{(Actual_{i}-Predicted_{i})^2}{t}, i=1,...,6,
$$ 
where $Actual_i$ are the actual economic variables at the $i$-th month, and $Predicted_i$ are the predicted ones simultaneously. 

The deep neural network models are trained for up to 100 epochs. The training process is terminated if MSE does not improve over 40 consecutive epochs. The final deep learning models (LSTM, GRU, and LSTM\_GRU together) were maintained at the lowest MSE.

\section{Applications} \label{sec3}
\begin{figure}[H]
\begin{center}
\includegraphics[scale = 0.4]{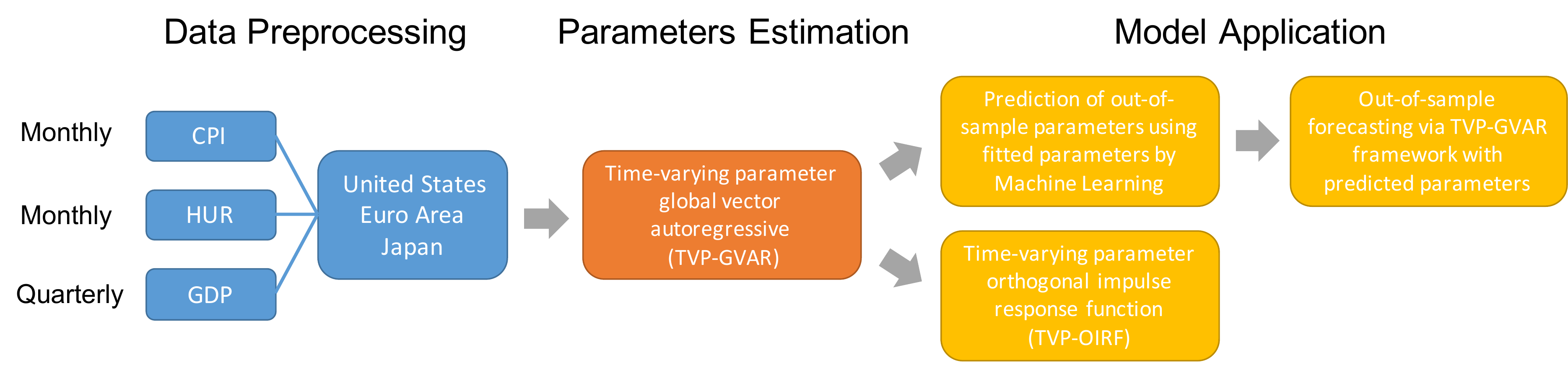}
\end{center}
\caption{\textbf{The workflow} }\label{workflow}
\end{figure}

\indent\indent We choose the United States, the Euro area, and Japan as the representative countries,  with the change in oil prices serving as the common shock. The desirable economic variables are CPI, HUR, and GDP with varied frequencies. Data is gathered on the first day of every month from the first of 2000 until the last of 2020. The comprehensive workflow is depicted in Figure \ref{workflow}. 

\begin{figure}[H]
\begin{center}
\includegraphics[scale = 0.5]{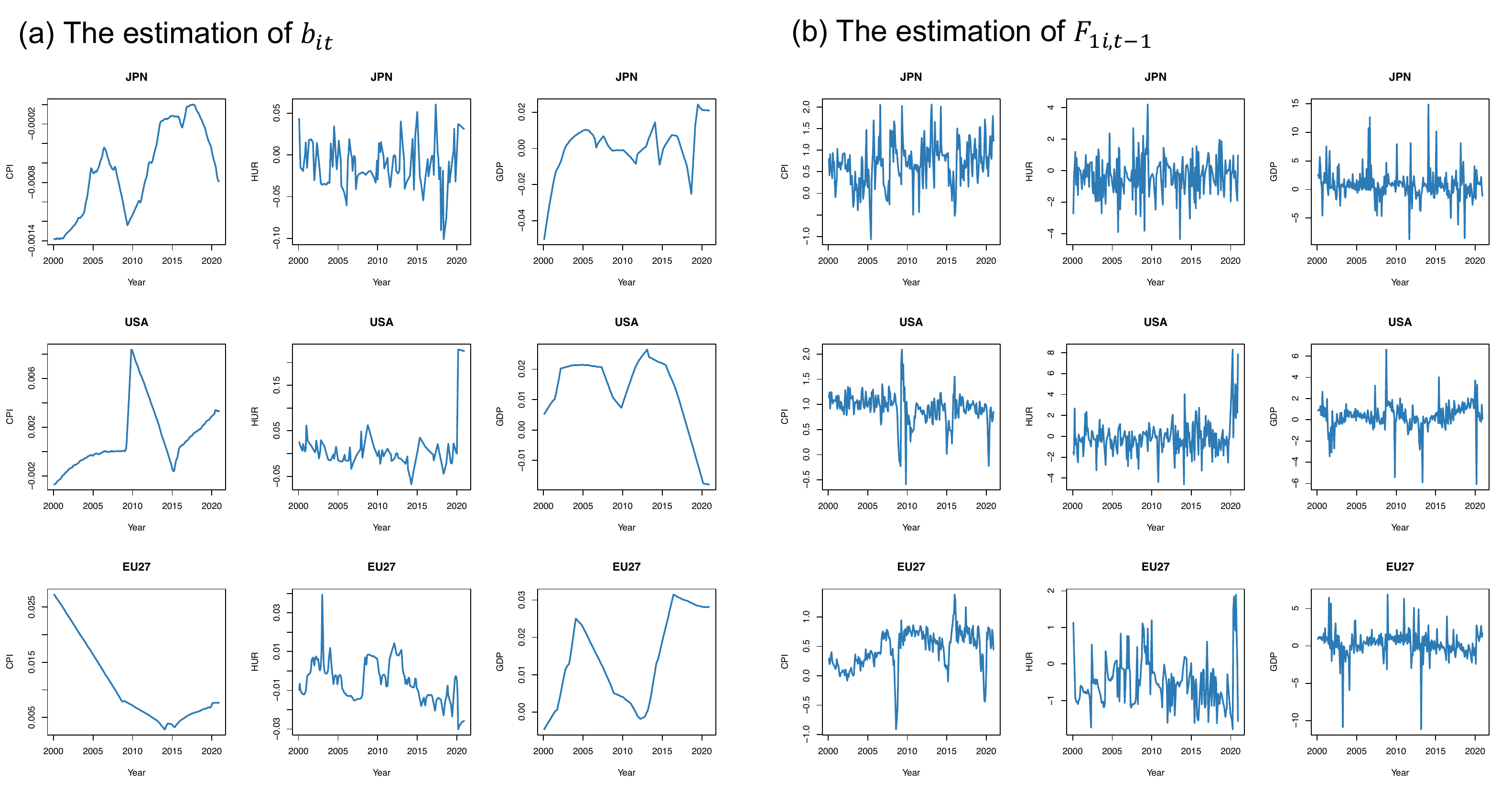}
\end{center}
\caption{\textbf{The trends of parameters for CPI, HUR, and GDP in the United States, Euro area, and Japan.} The (a) and (b) panels represent $b_{it}$ and $F_{1i,t-1}$. } \label{fig1}
\end{figure}

Firstly, we estimate the parameters of the time-varying GVAR model. The estimated parameters in Equation (\ref{equ:full}) for each output for each of the three countries are shown in Figure \ref{fig1}.
 
\begin{figure}[H]
\begin{center}
\includegraphics[scale = 0.5]{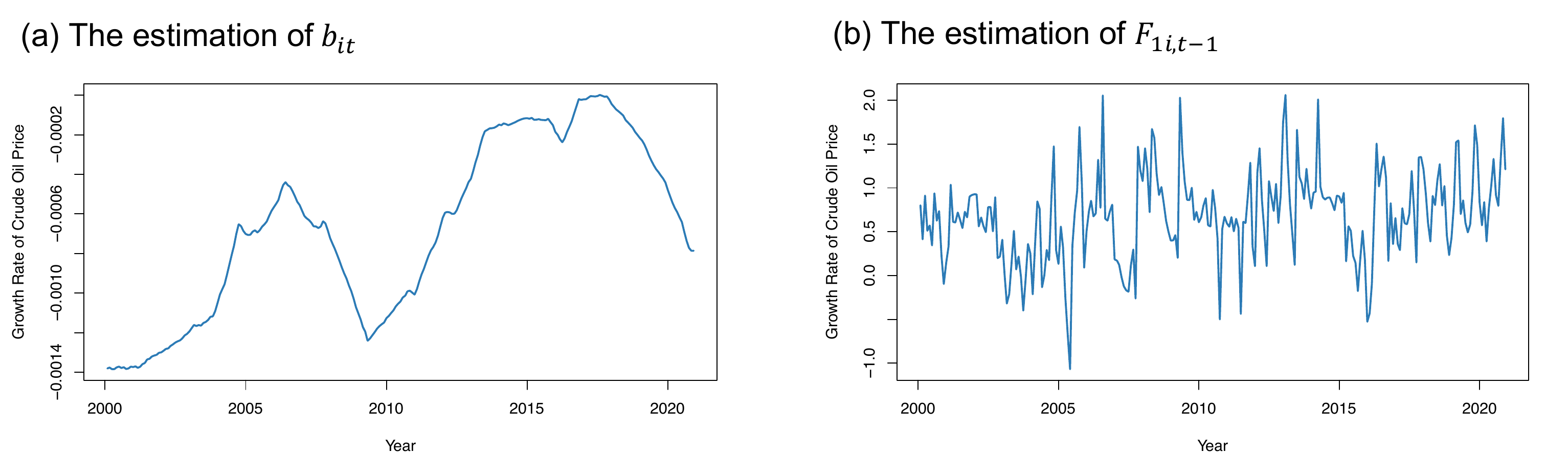}
\end{center}
\caption{\textbf{The trends of parameters for the oil price.} The (a) and (b) panels represent $b_{it}$ and $F_{1i,t-1}$. } \label{fig2}
\end{figure}

Figure \ref{fig2} shows the same parameters for the oil price. 
 
\begin{figure}[H]
\begin{center}
\begin{minipage}[t]{0.48\textwidth}
	\centering
	\footnotesize
	(a) The time-invariant OIRF
	\includegraphics[scale = 0.3]{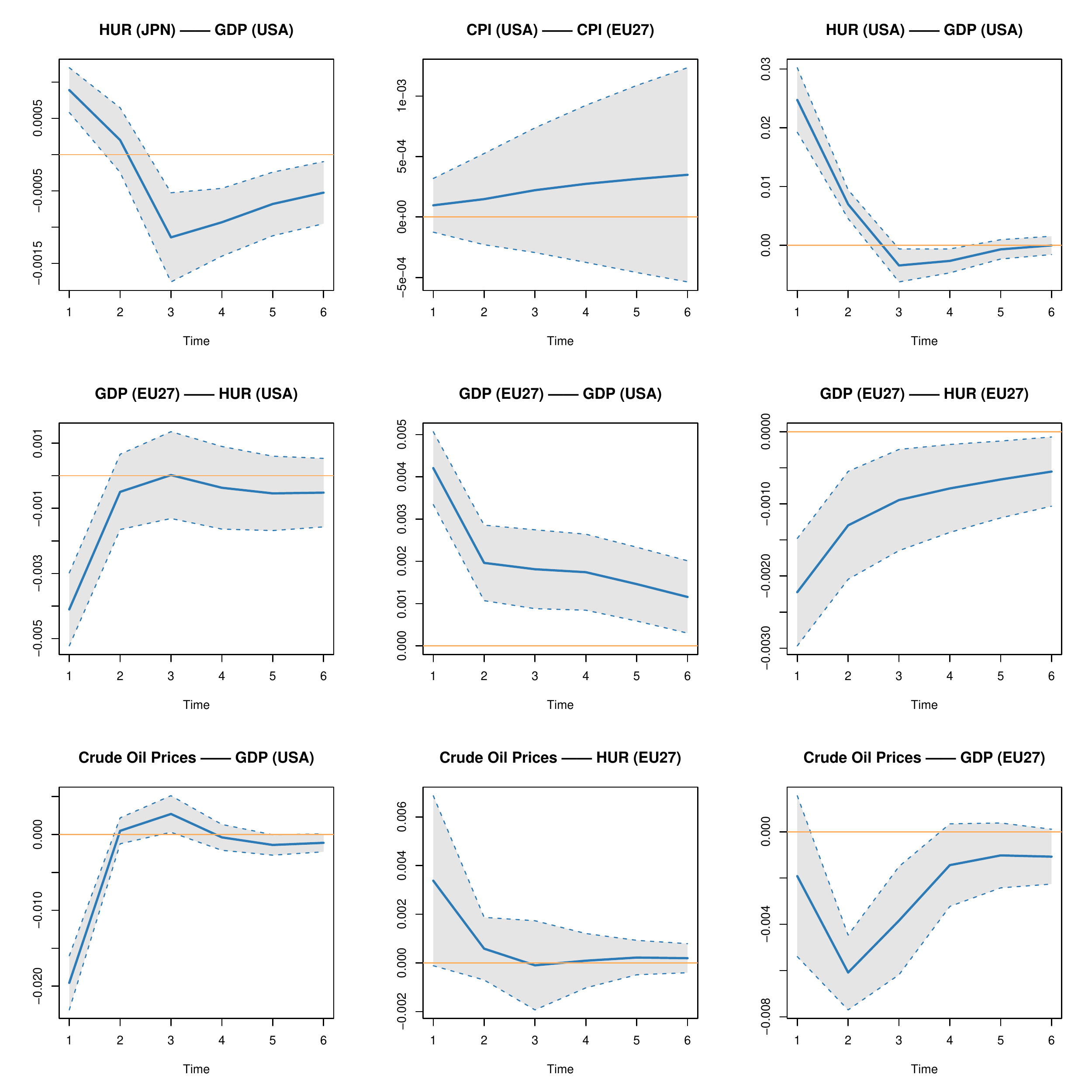}
\end{minipage}
\begin{minipage}[t]{0.48\textwidth}
	\centering
	\footnotesize
	(b) The time-varying OIRF of July 1st, 2020
	\includegraphics[scale = 0.3]{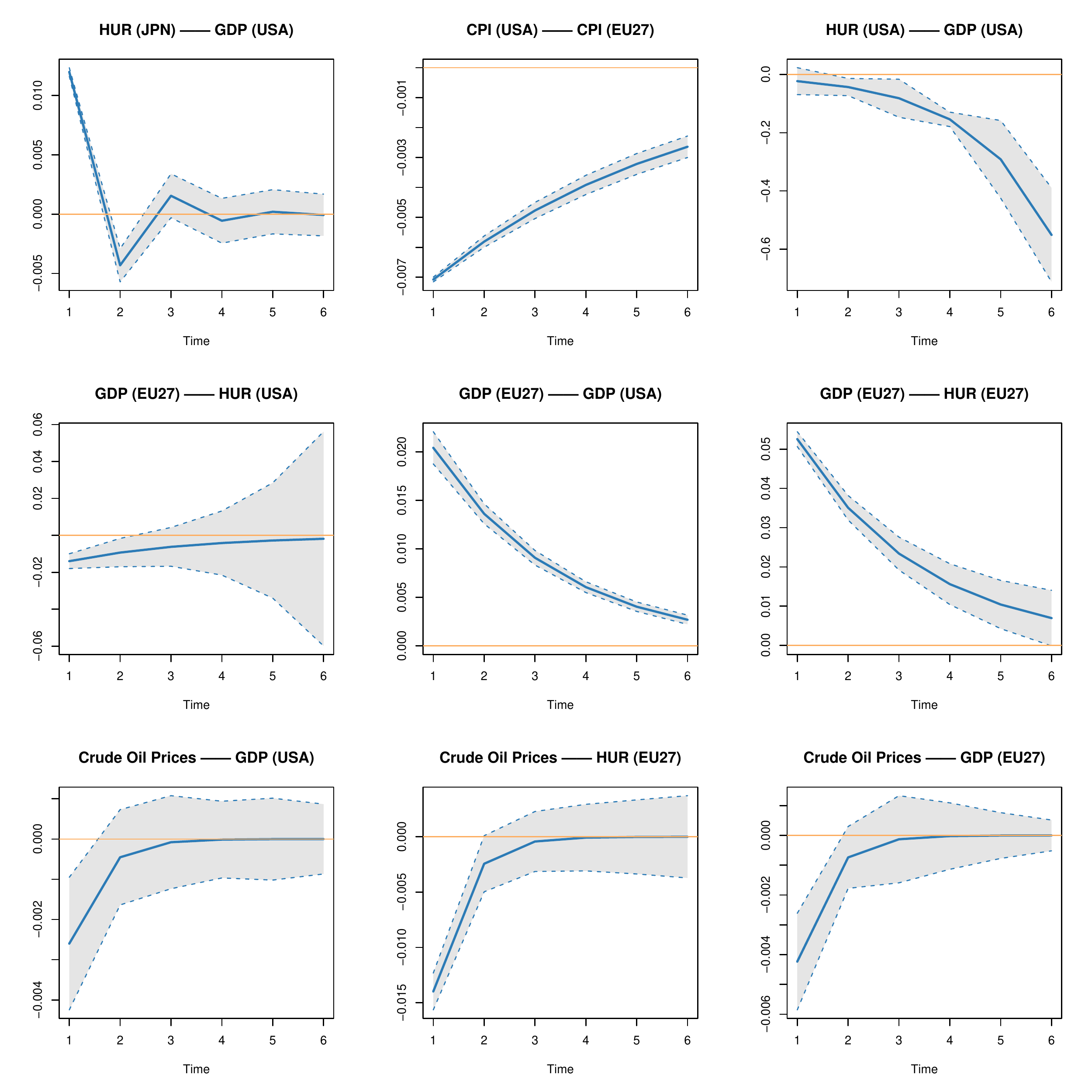}
\end{minipage}
\end{center}
\begin{center}
	\begin{minipage}[t]{0.48\textwidth}
		\centering
		\footnotesize
		The time-varying OIRF of April 1st, 2011
		\includegraphics[scale = 0.3]{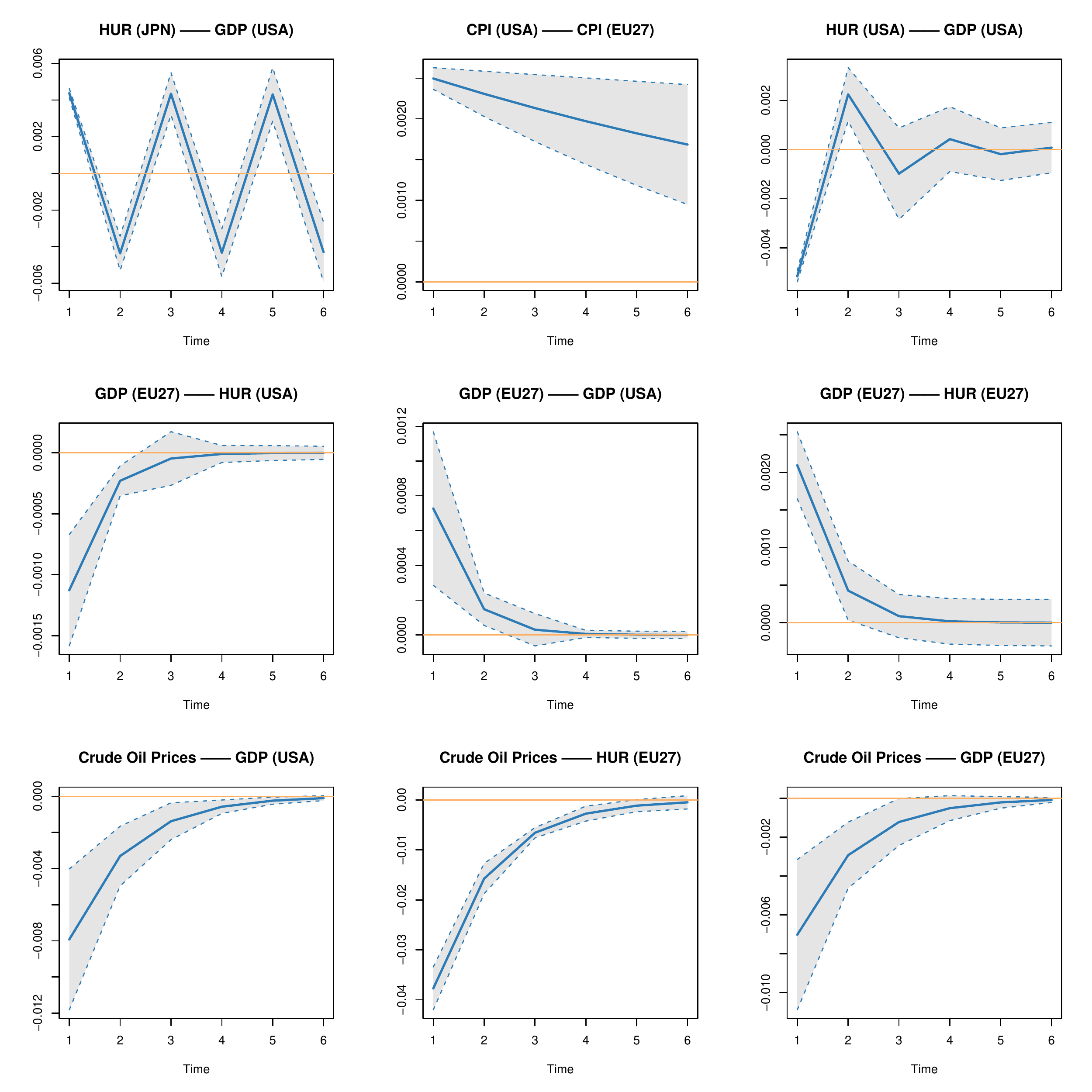}
	\end{minipage}
	\begin{minipage}[t]{0.48\textwidth}
		\centering
		\footnotesize
		(d) The time-varying OIRF of December 1st, 2007
		\includegraphics[scale = 0.3]{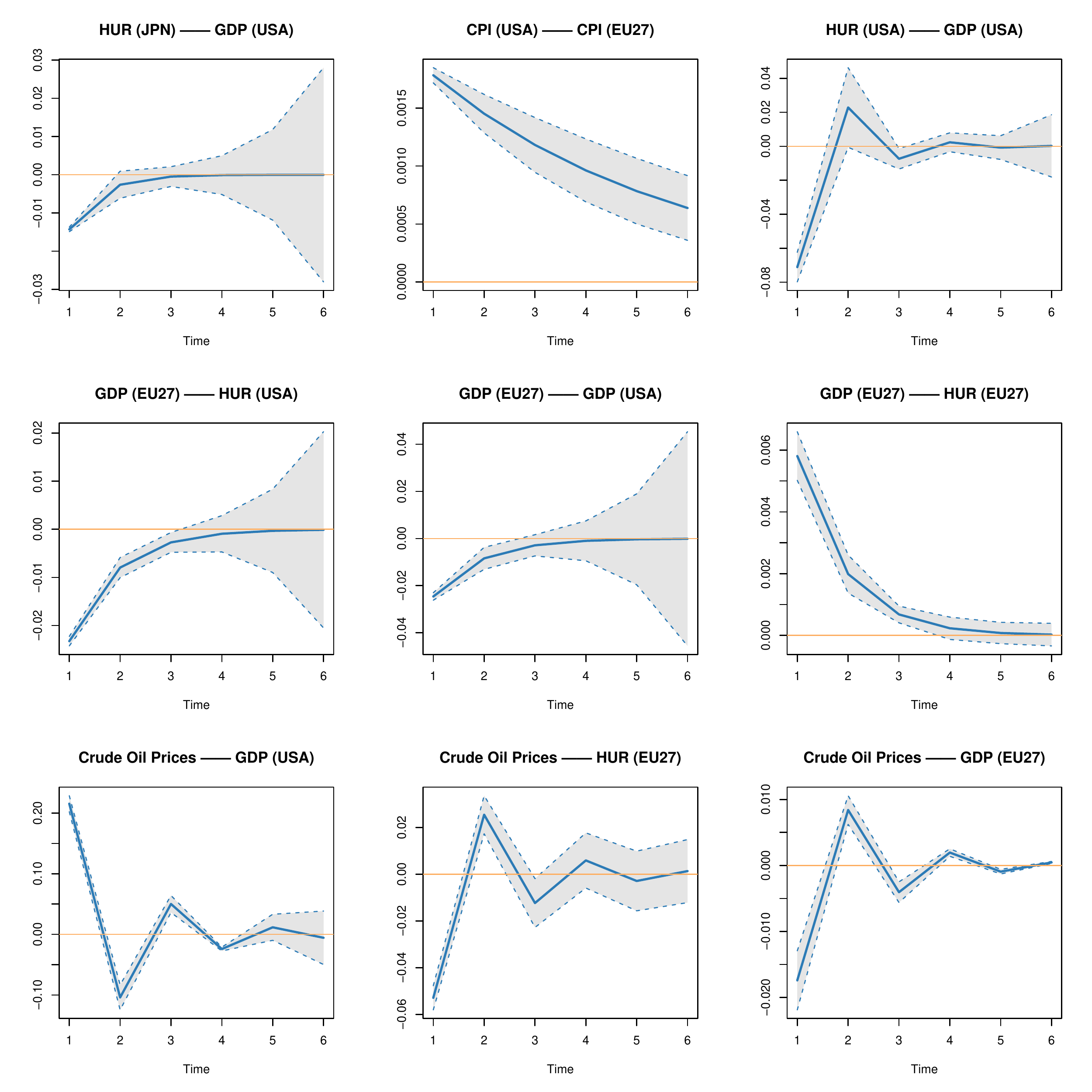}
	\end{minipage}
\end{center}
\caption{\textbf{The time-invariant OIRF and time-varying OIRF of three different time points with asymptotic bounds}. The figure displays the mean orthogonal impulse responses (solid blue lines) of partial economic variables to a unit shock. Each subgraph shows how the variables fluctuate over the subsequent six periods after a unit oil price shock of another economic variable in a region. The shaded areas surrounding the impulse responses represent 95\% asymptotic bounds. The solid red lines correspond to 0.} \label{fig3}
\end{figure}

Secondly, the OIRF and its asymptotic confidence intervals can be estimated using the acquired GVAR parameters and the asymptotic distribution of OIRF. Figure \ref{fig3} describes the partial findings of the time-invariant OIRF and time-varying OIRF of three different time points for desirable economic variables in the United States, Euro area, and Japan and their asymptotic bounds. Specifically, we investigate three-time issues, July 1st, 2020, April 1st, 2011, and December 1st, 2007, corresponding to the global COVID-19 confirmed cases of over 10 million, the Fukushima nuclear meltdown in Japan, and the sub-prime mortgage crisis in the United States, respectively.

After a predetermined amount of time, as shown in Figure \ref{fig3}, practically all OIRF will converge to the value 0. Here, we focus on the upcoming period of six months, except for the time-varying OIRF of the GDP of the United States to the unit shock from the HUR of it on July 1st, 2020. It reveals that these economic variables are a reasonable degree of consistency in this GVAR system, regardless of whether its parameters are time-varying. Additionally, a positive sign on the OIRF indicates a positive association between the shock and response and vice versa. For instance, we can see from the first subgraph of the second row of Figure \ref{fig3}(a) that following a unit shock from the Euro area's GDP, the HUR of the United States decreases, and the impact lessens dramatically to around 0 a month later. In general, the graphic demonstrates that the effects represented by OIRF vary between the time-invariant and time-varying GVAR models and within different time points of the time-varying GVAR model. As an illustration, consider the OIRF of the GDP of the United States to the unit shock from the Euro area's GDP, which is the central subgraph of each panel in Figure \ref{fig3}. The impulse responses of the time-varying and time-invariant systems were positive on July 1st, 2020, and April 1st, 2011. However, the time-varying model's outcome of December 1st, 2007, was negative.

To go into more detail, according to the characteristics of IRF, we can draw certain practical implications in both the time-invariant and time-variant GVAR model. On the one hand, in the time-invariant mode, it can be seen from the third subgraph of the second row of Figure \ref{fig3}(a) reveals that the HUR of the Euro area experiences a decline after a unit shock originating from its GDP. However, the impact gradually lessens until it is nonexistent as more time passes, resulting in the expansion of the GDP being beneficial to lowering the HUR in the typical scenarios in Europe. In addition, the GDP of the United States experiences an increase at the same time that the GDP of the Euro area grows by one unit. Nevertheless, this impact is still significant even six months after the fact, as shown by the central subgraph of Figure \ref{fig3}(a), indicating the notable economic ties between nations on different continents.
On the other hand, when it comes to the time-invariant OIRF, it can be seen in the second subgraph of the last row of Figure \ref{fig3}(b) that the response of the Euro area's HUR to the shock from oil prices became negative after the COVID-19 confirmed cases of over 10 million. At the same time, the same effect is positive in the time-invariant model. This illustrates that Europe's economic performance is anomalous during global health emergencies. Secondly, in April 2011, when the Fukushima nuclear meltdown took place (Figure \ref{fig3}(c)), the unit shock from Japan's HUR caused an abnormal shaking of the GDP of the United States, showing that the influence generated by the Japanese economy was unstable during that specific period. Finally, it is represented by the first subgraph of the last row of Figure \ref{fig3}(d) that the response of the GDP of the United States to the unit shock from oil prices was positive on December 1st, 2007, which is contrary to all other time points and the time-invariant model. The relationship between the United States and the world economy changed during the sub-prime mortgage crisis.

Due to the characteristics of OIRF, its reaction to shocks from several factors can be constructed by summing the responses of a single variable to shocks from a number of other economic variables.  For example, Figure \ref{fig3} shows the correlations between multiple shocks and the impulse responses of the US GDP and the HUR of the Euro area, respectively. Figure \ref{fig:OIRF_n_to_1} is the result of calculating the OIRF summary for the same output.

\begin{figure}[H]
	\begin{center}
		\begin{minipage}[t]{0.48\textwidth}
			\centering
			\footnotesize
			(a1) The time-invariant OIRF
			\includegraphics[scale = 0.4]{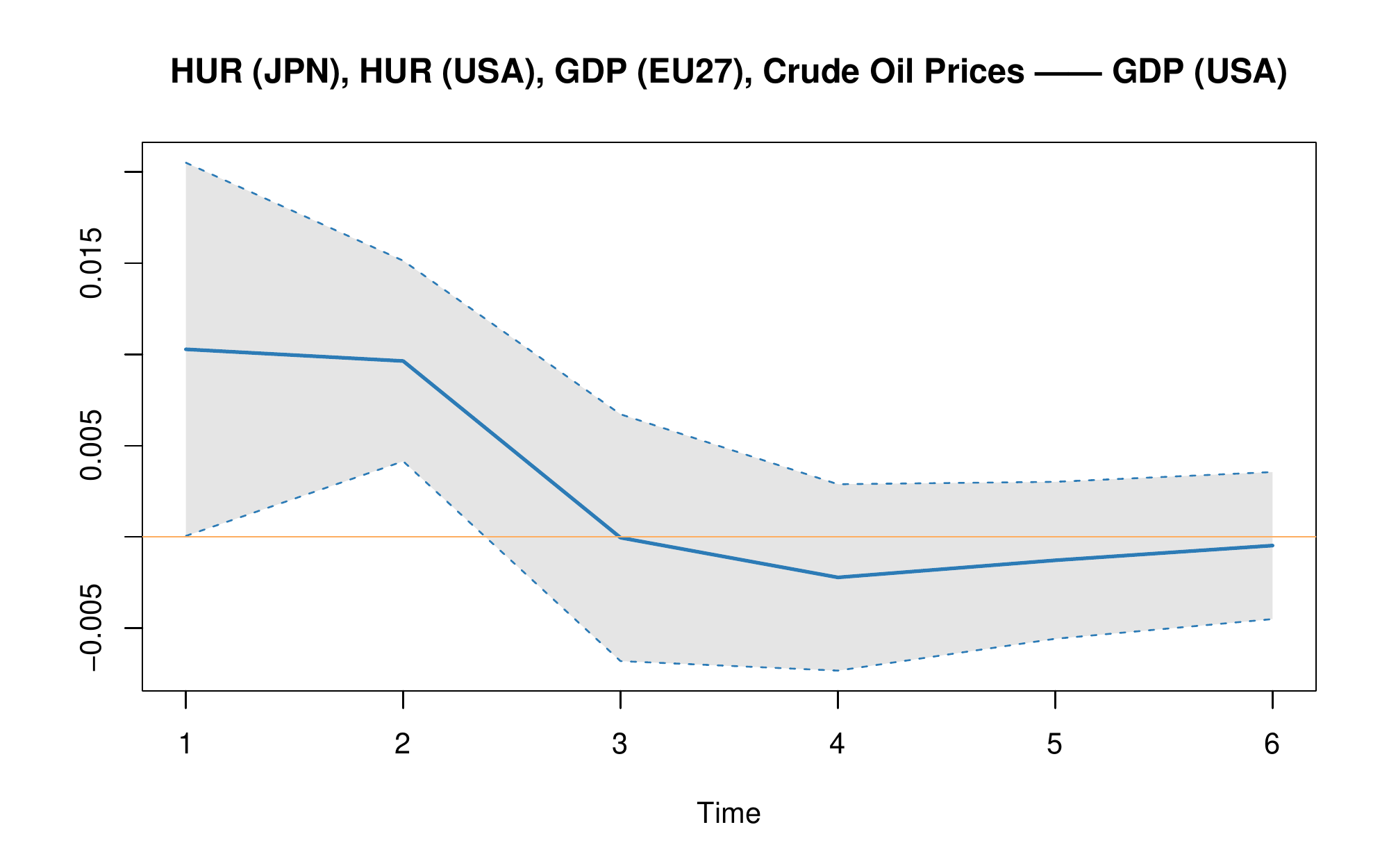}
		\end{minipage}
		\begin{minipage}[t]{0.48\textwidth}
			\centering
			\footnotesize
			(b1) The time-varying OIRF of July 1st, 2020
			\includegraphics[scale = 0.4]{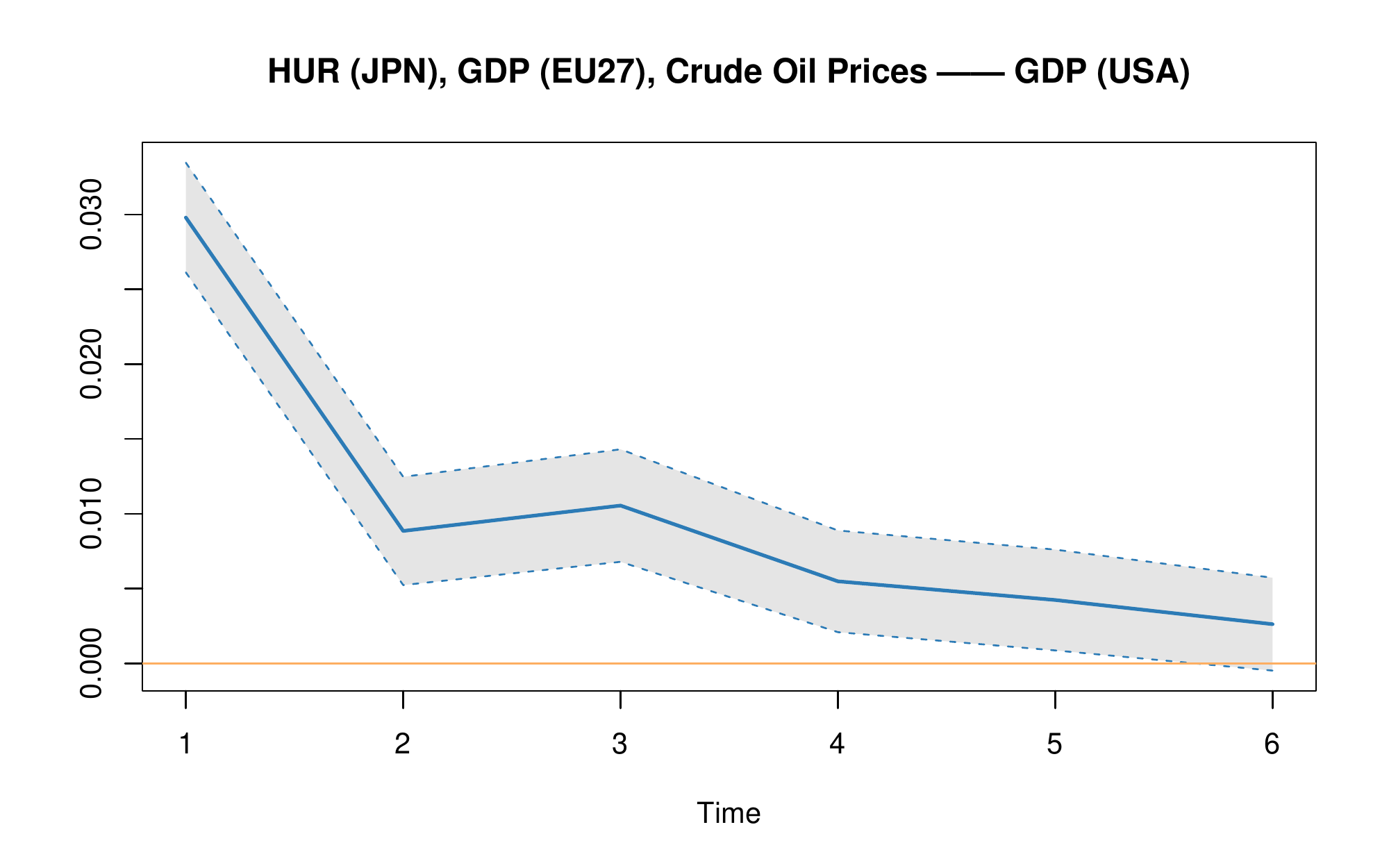}
		\end{minipage}
	\end{center}
	\begin{center}
		\begin{minipage}[t]{0.48\textwidth}
			\centering
			\footnotesize
			(c1) The time-varying OIRF of April 1st, 2011
			\includegraphics[scale = 0.4]{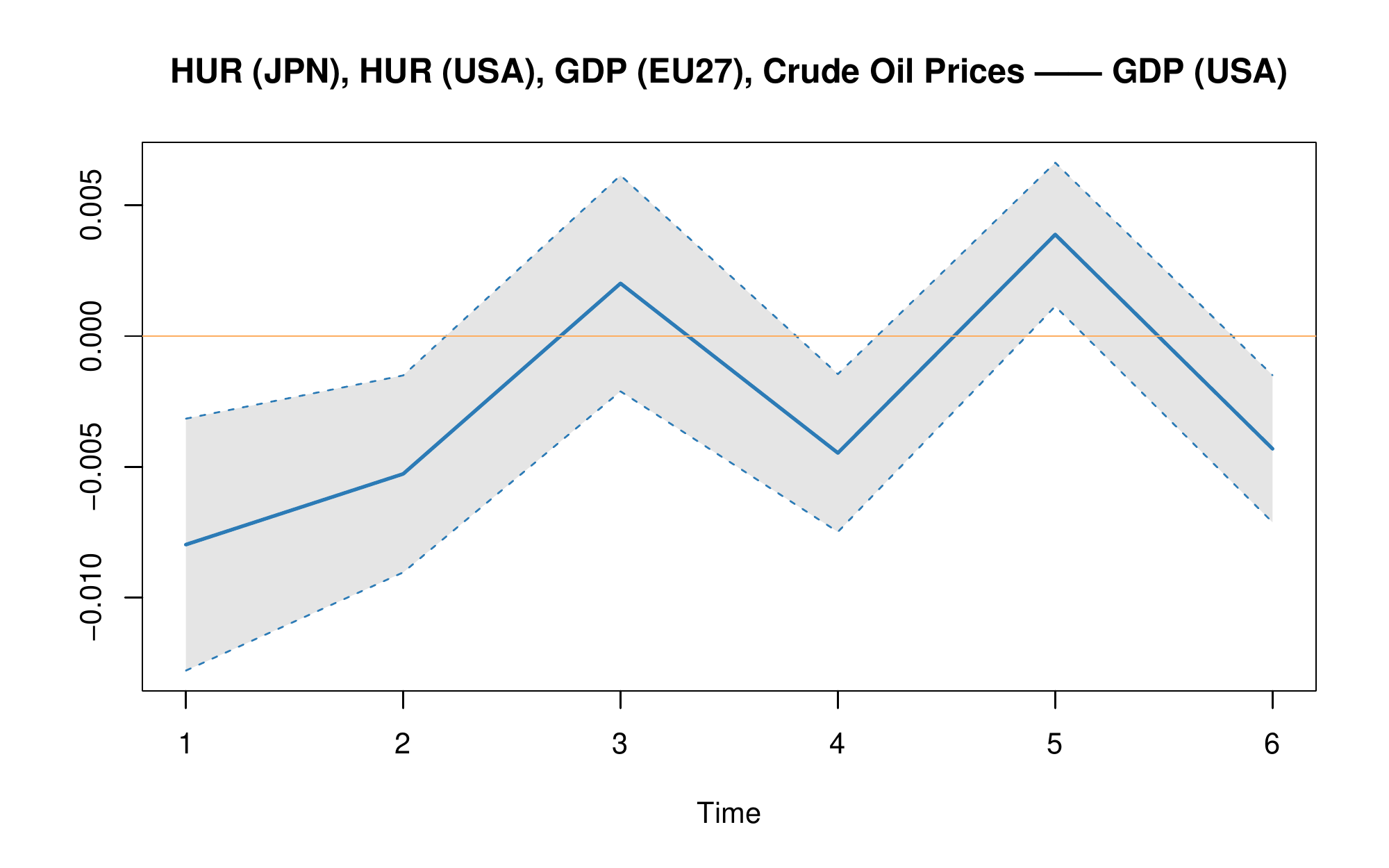}
		\end{minipage}
		\begin{minipage}[t]{0.48\textwidth}
			\centering
			\footnotesize
			(d1) The time-varying OIRF of December 1st, 2007
			\includegraphics[scale = 0.4]{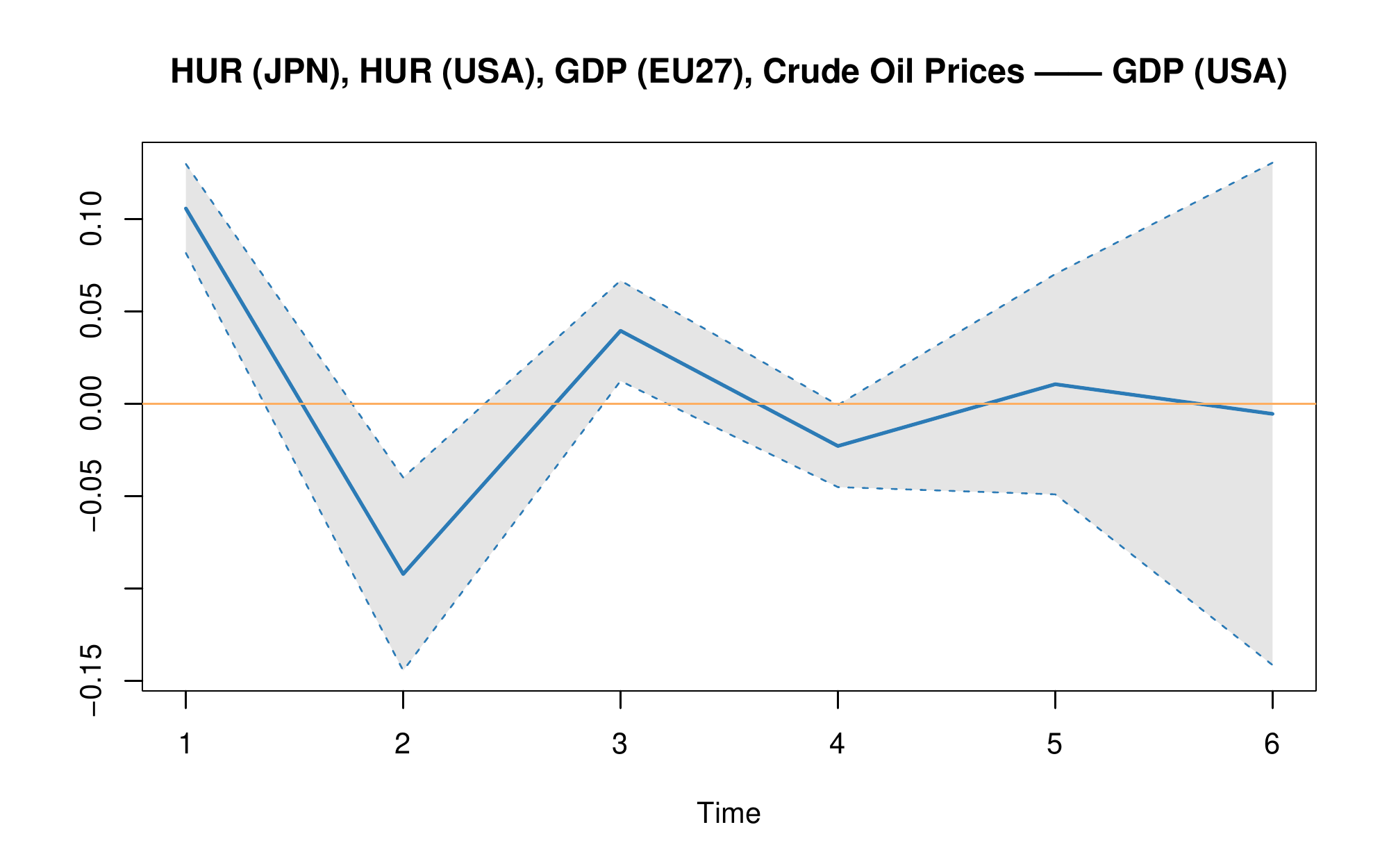}
		\end{minipage}
	\end{center}
	\begin{center}
	\begin{minipage}[t]{0.48\textwidth}
		\centering
		\footnotesize
		(a2) The time-invariant OIRF
		\includegraphics[scale = 0.4]{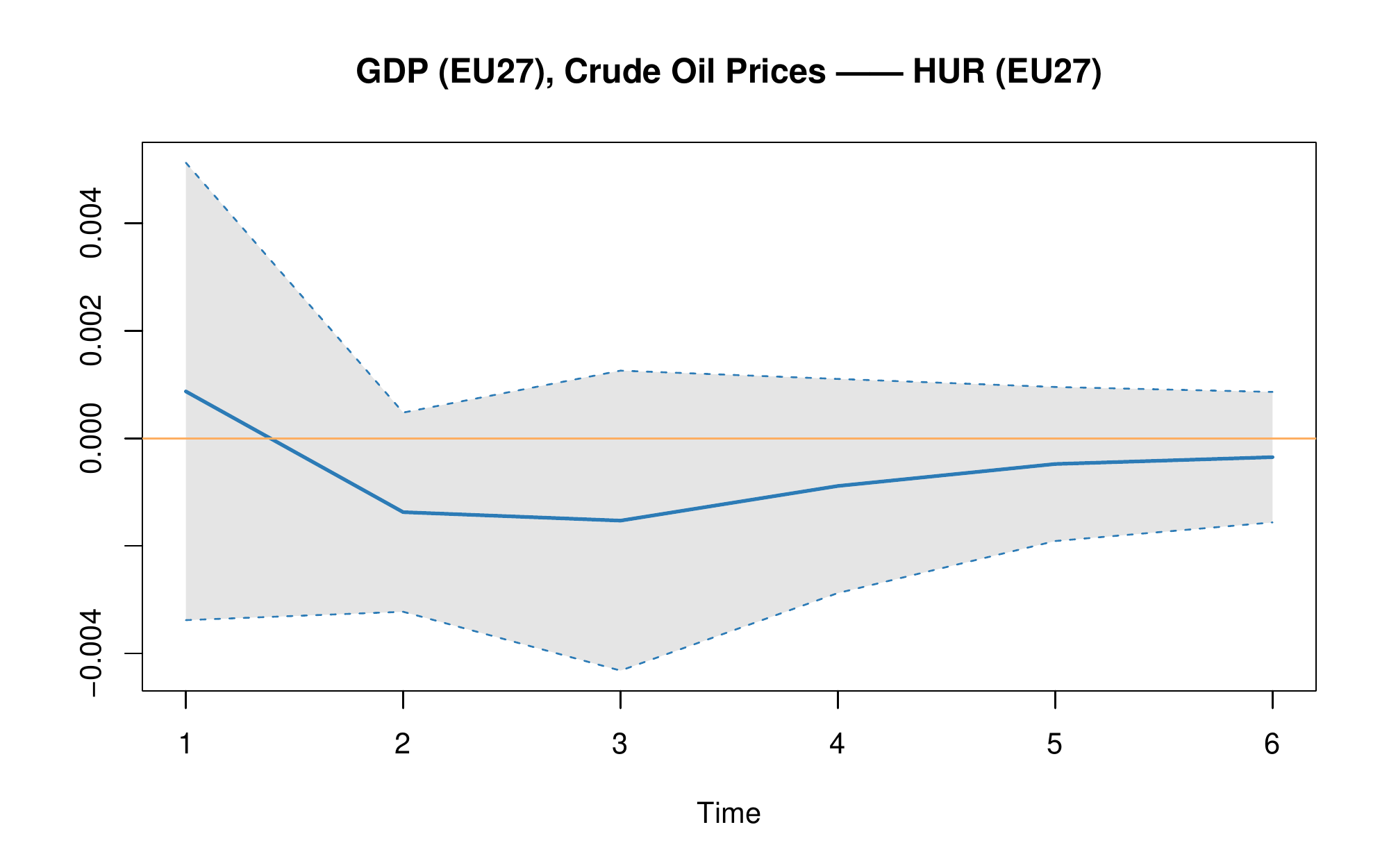}
	\end{minipage}
	\begin{minipage}[t]{0.48\textwidth}
		\centering
		\footnotesize
		(b2) The time-varying OIRF of July 1st, 2020
		\includegraphics[scale = 0.4]{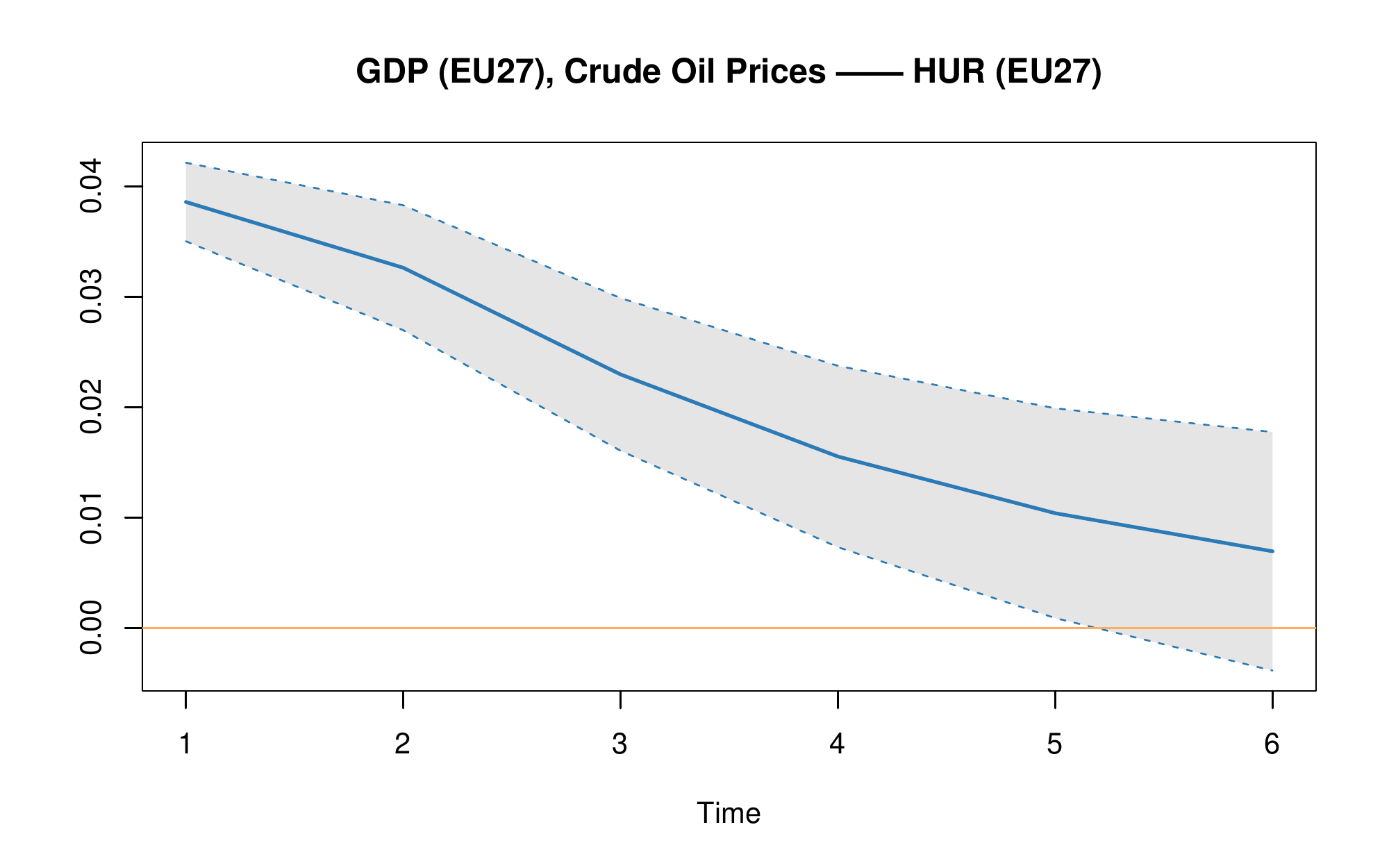}
	\end{minipage}
\end{center}
\begin{center}
	\begin{minipage}[t]{0.48\textwidth}
		\centering
		\footnotesize
		(c2) The time-varying OIRF of April 1st, 2011
		\includegraphics[scale = 0.4]{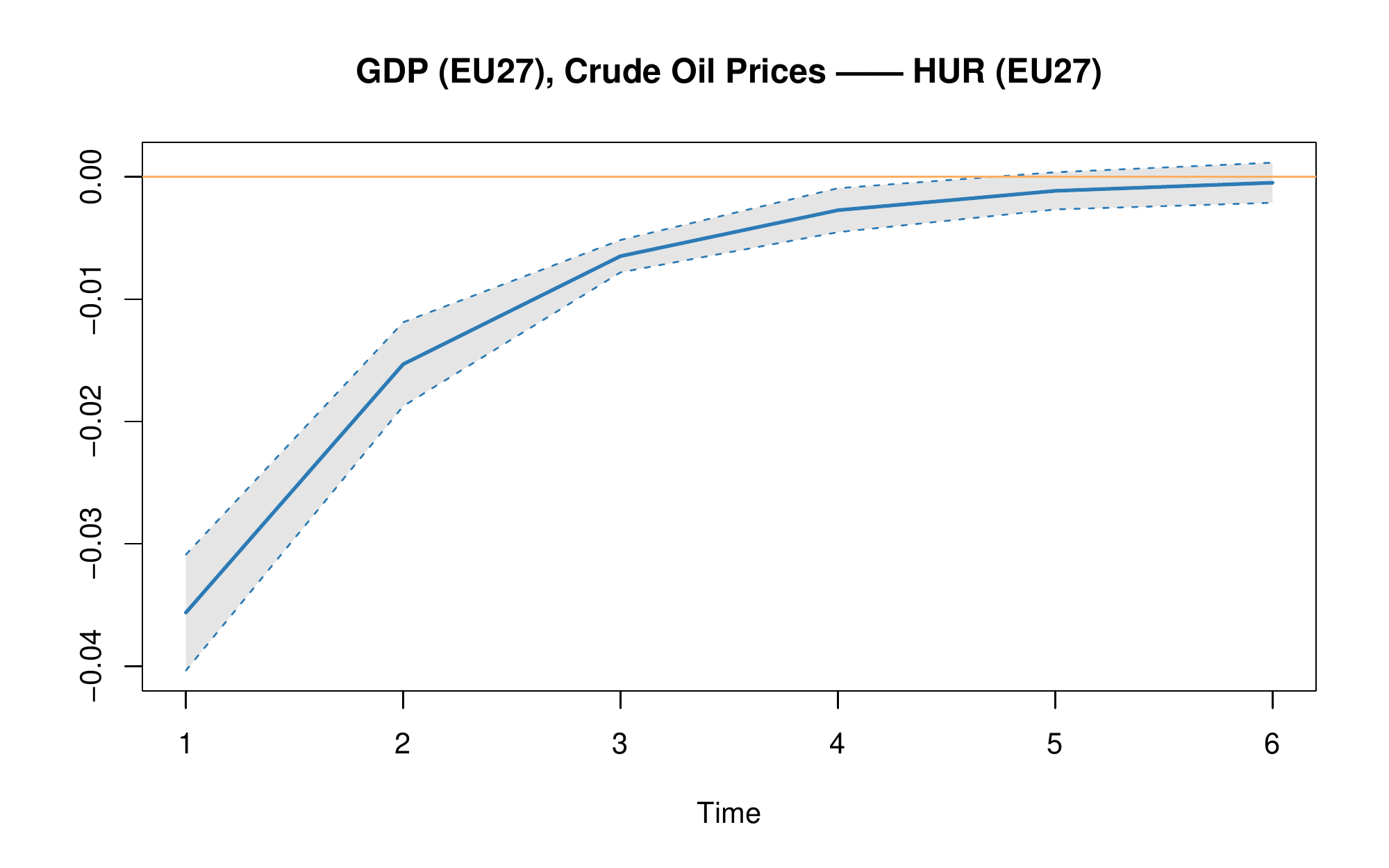}
	\end{minipage}
	\begin{minipage}[t]{0.48\textwidth}
		\centering
		\footnotesize
		(d2) The time-varying OIRF of December 1st, 2007
		\includegraphics[scale = 0.4]{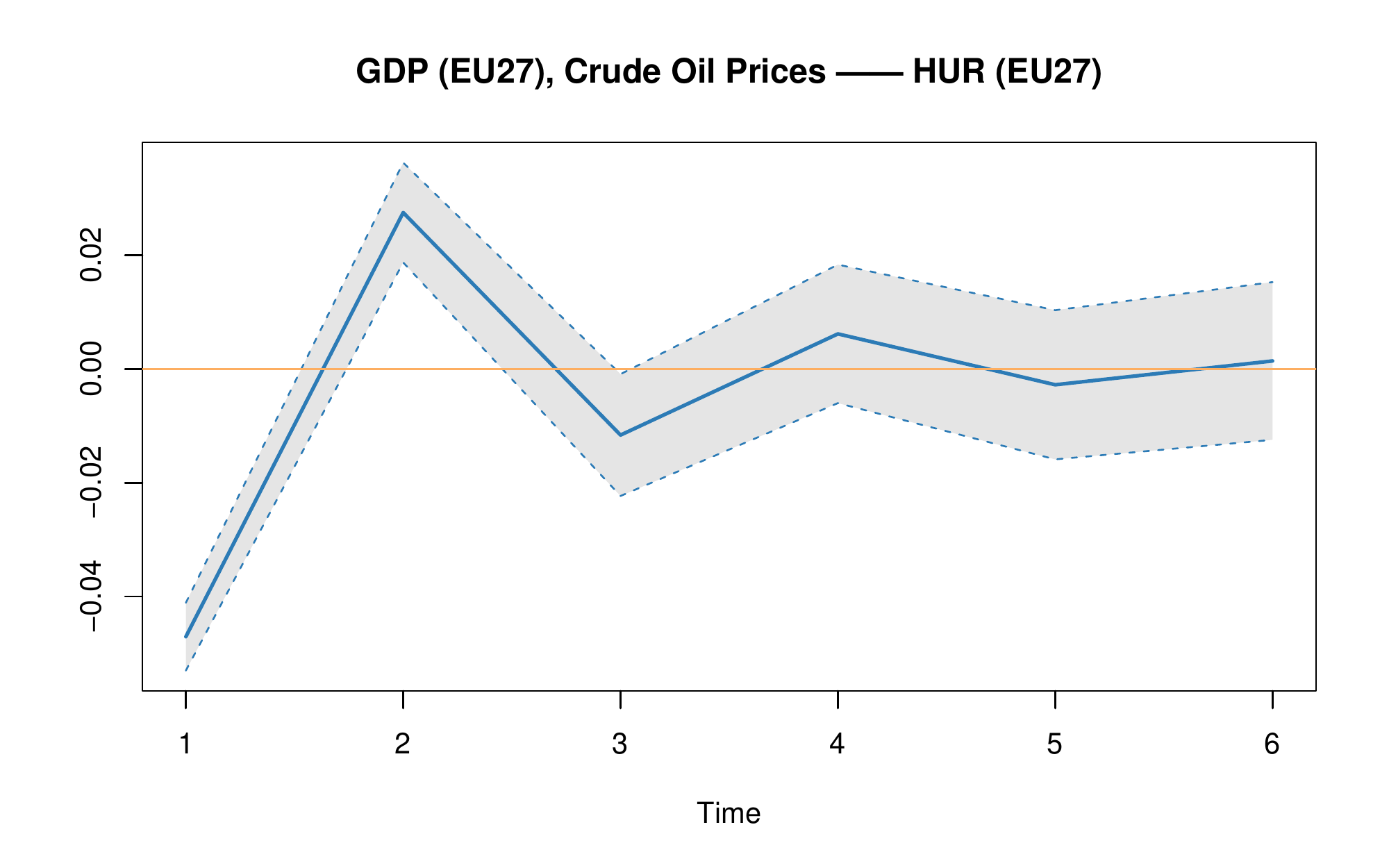}
	\end{minipage}
\end{center}
\end{figure}
\begin{figure}
	\caption{\textbf{The time-invariant and time-varying OIRF to multivariate shocks with asymptotic bounds}. The figure displays the mean impulse responses (solid blue lines) of the United States GDP and the Euro area's HUR shocks from several other economic variables. Each subgraph shows how these two variables fluctuate over the subsequent six periods after the shocks in a region. The shaded areas surrounding the impulse responses represent 95\% asymptotic bounds. The solid red lines correspond to 0.} \label{fig:OIRF_n_to_1}
\end{figure} 

Since many of the shapes of the OIRF to multivariate shocks are similar to one of the OIRF to univariate shock, it can be inferred from Figure \ref{fig:OIRF_n_to_1} that the OIRF to shocks from more than one output is frequently dominated by one particular variable. For example, on April 1st, 2011, and December 1st,  2007, the OIRF of the Euro area's HUR to the shocks from its GDP and oil prices (Figure \ref{fig:OIRF_n_to_1}(a2)(c2)(d2)) is quite similar to the OIRF to the shock from oil prices alone (the second subgraph of the last row of Figure \ref{fig3}(a)(c)(d)). This is the case for both the time-varying GVAR and time-invariant models. Thus, during these two corresponding economic crises and shared situations, the shock from oil prices had a much more significant impact on the Euro area's HUR than the effect on its GDP. Nevertheless, on July 1st, 2020, of the time-varying system, the same OIRF (Figure \ref{fig:OIRF_n_to_1}(b2)) now looks like the OIRF to the shock only from the Euro area's GDP (the third subgraph of the second row of Figure \ref{fig3}(b)).

\begin{figure}[H]
\begin{center}
\includegraphics[scale = 0.5]{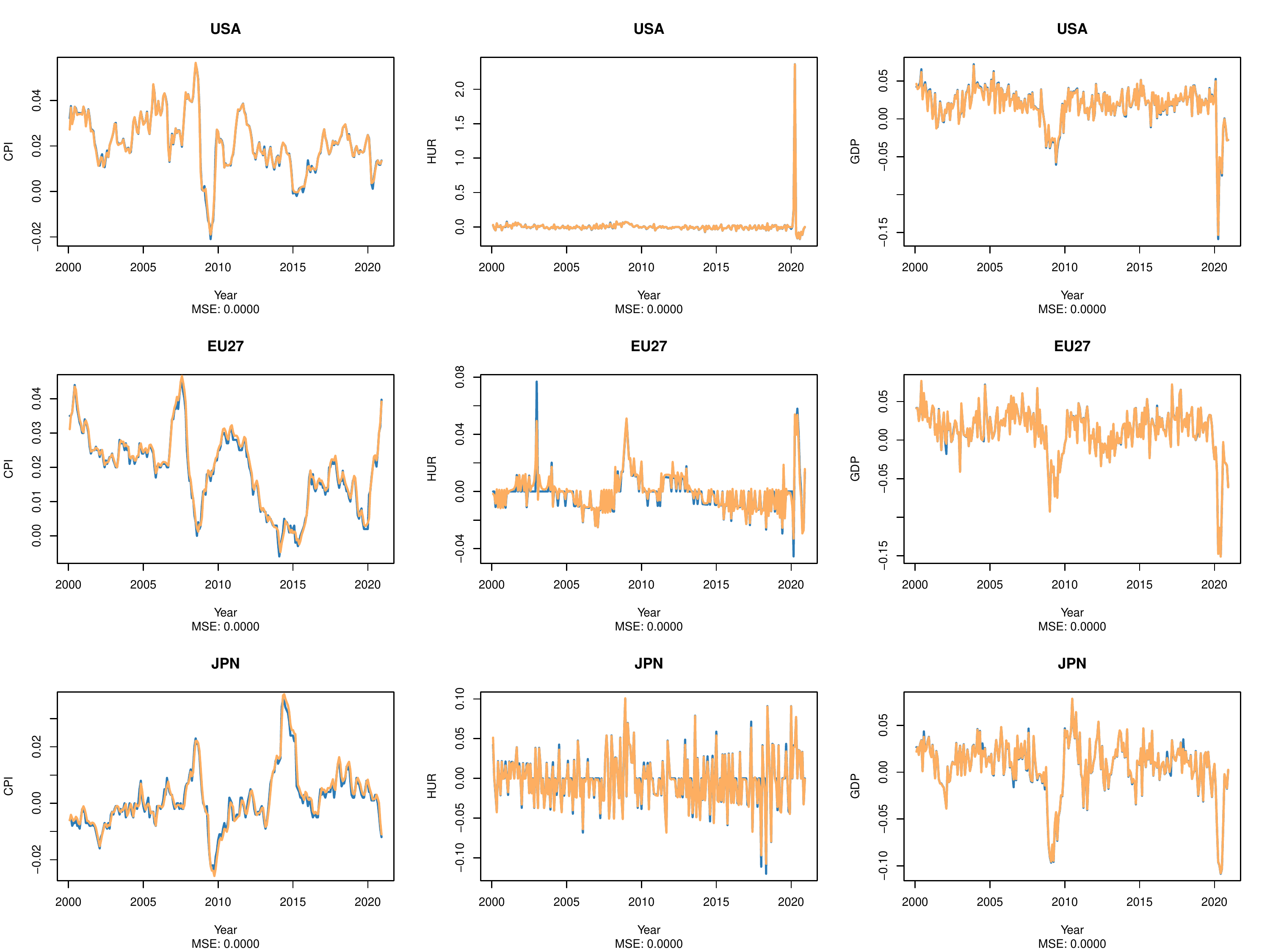}
\end{center}
\caption{\textbf{The predicted and the observed CPI, HUR, and GDP in the United States, Euro area, and Japan based on the samples.}} \label{fig4}
\end{figure}

\begin{figure}[H]
\begin{center}
\includegraphics[scale = 0.5]{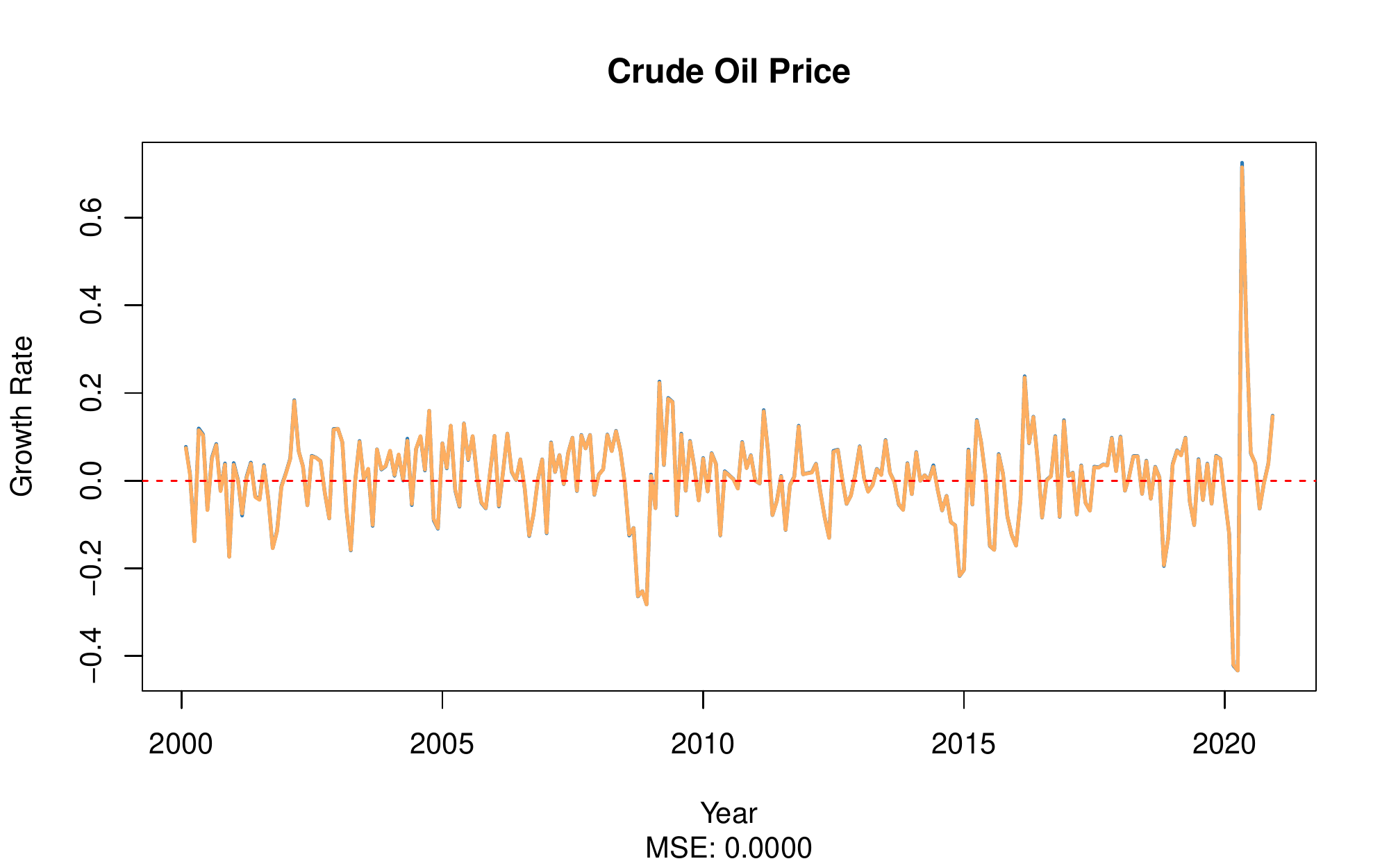}
\end{center}
\caption{\textbf{The predicted and the observed oil price based on the samples.}} \label{fig5}
\end{figure}

\begin{figure}[H]
\begin{center}
\includegraphics[scale = 0.5]{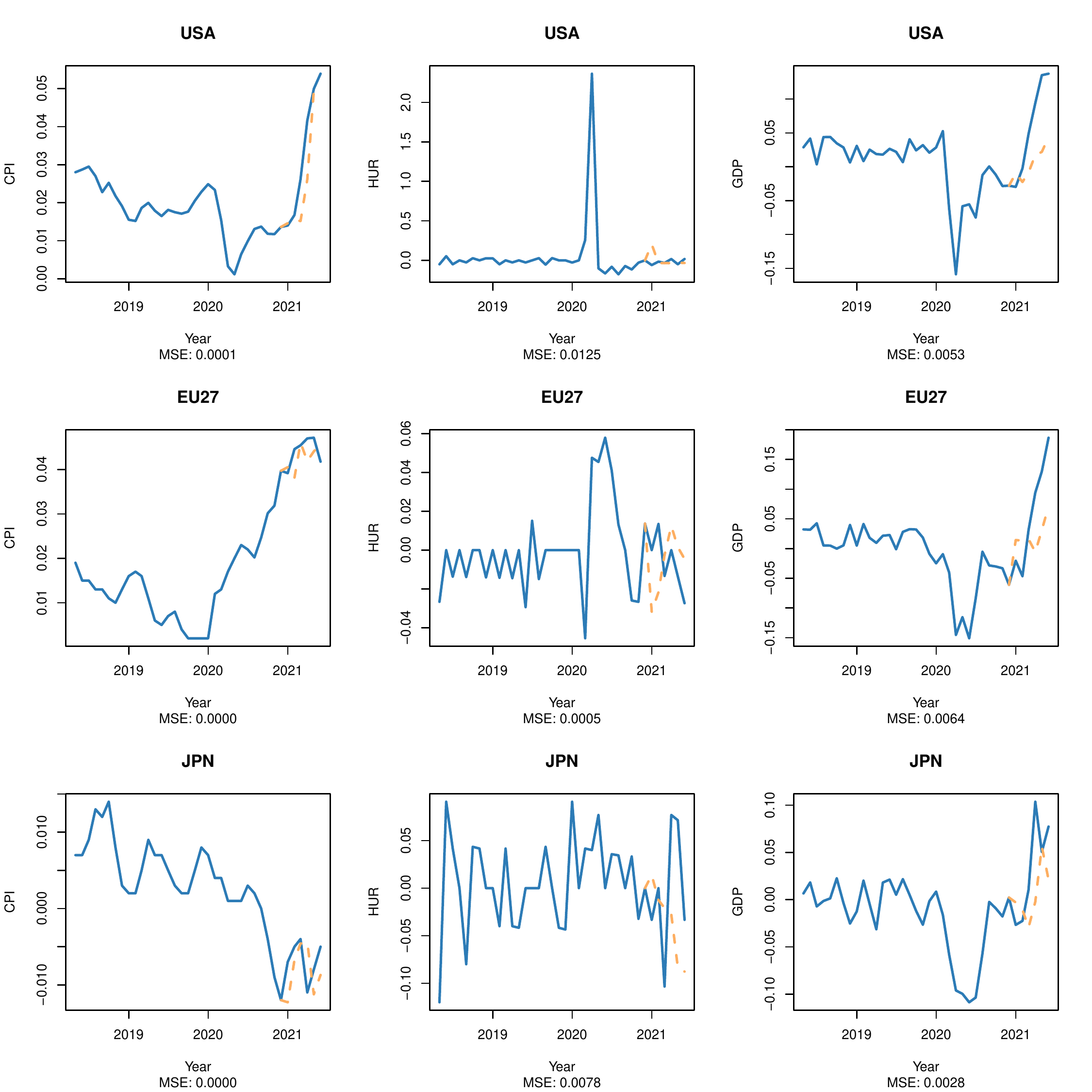}
\end{center}
\caption{\textbf{The predicted and the observed CPI, HUR, and GDP in the United States, Euro area, and Japan for the following six months (LASSO).}} \label{fig6}
\end{figure}

\begin{figure}[H]
\begin{center}
\includegraphics[scale = 0.5]{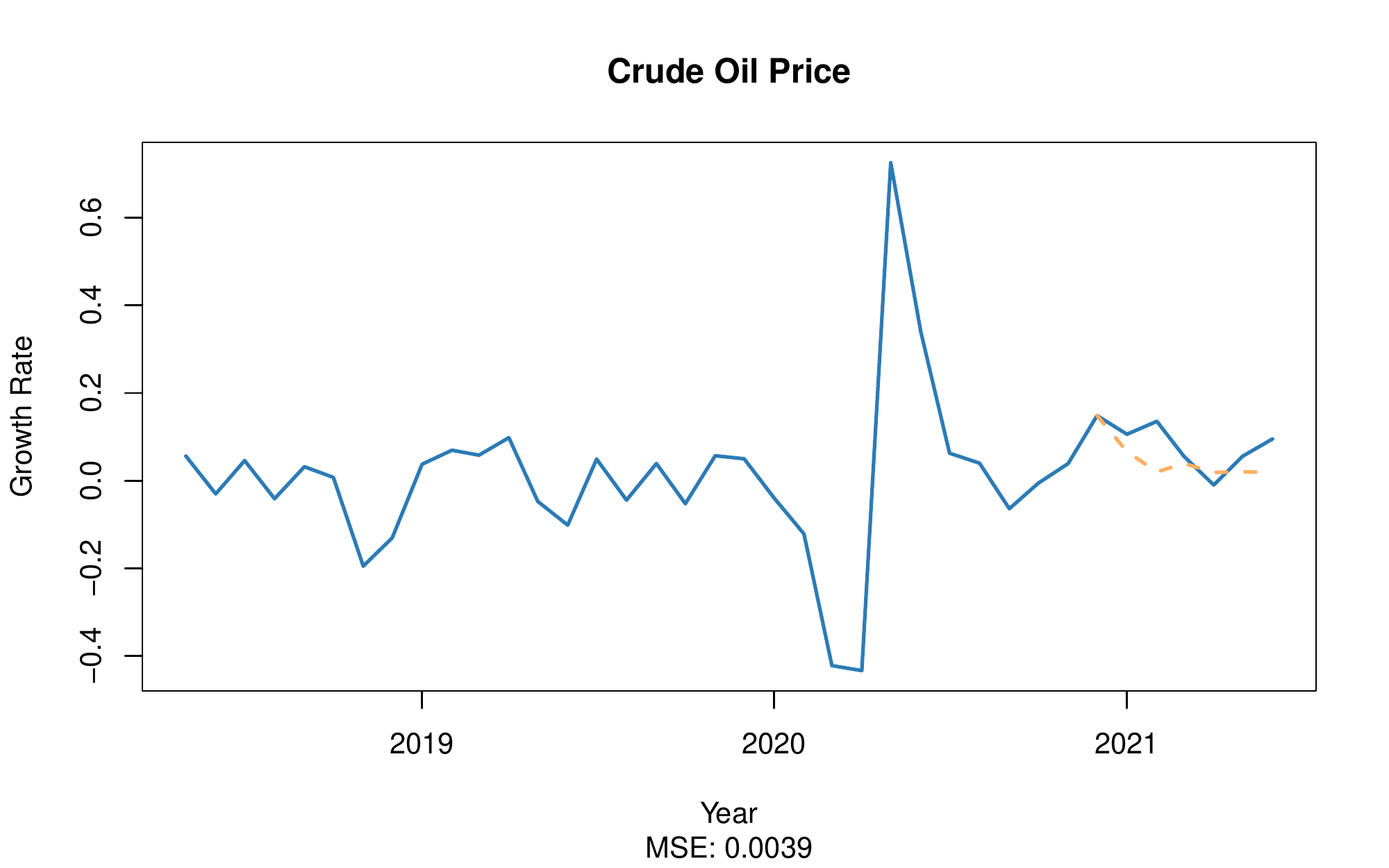}
\end{center}
\caption{\textbf{The predicted and the observed oil price for the following six months (LASSO).}} \label{fig7}
\end{figure}

\begin{table}[H]
\center
\begin{tabular}{lclc} 
\hline
Methods&MSE&Methods&MSE\\
\hline
CONSTANT &	0.0446 &LASSO	&\textbf{0.0039}\\
VAR	&0.0119&LSTM	&0.0063\\
TREE&	0.0097&GRU	&0.0087\\
RF	&0.0088&LSTM\_GRU	&0.0112 \\
\hline
\end{tabular}
\caption{The values of MSE for each existing model.}
\label{tab1}
\end{table}
The MSE calculates the comparisons across all models (Table \ref{tab1}). In light of the information shown in Table \ref{tab1}, we use LASSO as our projection method for the economic variables and oil price because it has the lowest value of MSE. Figure \ref{fig4} presents the predicted and the observed CPI, HUR, and GDP in the United States, Euro area, and Japan. Again, the predicted values are incredibly close to the observed ones for each country. Thus, the predictions of our TVPs based on the samples from 2000 to 2020 are remarkably accurate. Moreover, it is comparable to how oil prices have behaved (Figure \ref{fig5}). Using our TVP-GVAR-ML, we have projected the following economic variables for six months beginning in January 2021 and ending in June 2021. Figure \ref{fig6} depicts the trajectories of CPI, HUR, and GDP , and the corresponding smaller MSEs imply that our models incorporating the LASSO-type technique have higher accuracy. This is similar to the predicted oil price in the following six months (Figure \ref{fig7}).       

\section{Conclusions} \label{sec4}
\indent\indent In our paper, we develop the TVP-GVAR merged with a machine learning approach for predicting the economic variables in developed regions. The results of model selection are reported based on the values of MSE. Both the in-sample and the out-of-sample analyses benefit from the relatively high precision brought about by the association of the TVP-GVAR estimate and prediction with the LASSO-type technique. Furthermore, we make predictions on the structure of the parameters, which enables high accuracy in time-varying out-of-sample projections and may reflect the potential influencing factors and temporal mechanisms. In this context, the chosen LASSO-type technique helps to lessen the likelihood of over-fitting, broadens the proposed model's applicability, and, as a result of its orthogonality, improves the accuracy of test samples. In addition, it is adaptable enough to integrate a wide variety of machine learning and network structures, resulting in significantly lower values of MSE. This data-driven procedure is easily accessible for practitioners in pertinent application scenarios.

We also consider time-varying and time-invariant orthogonal impulse response functions and derive the asymptotic distribution of estimates on bounds of OIRF, demonstrating the influence on each variable after a unit shock from every other variable. In practice, the estimated OIRF shows that most of these economic variables are generally stable in the GVAR system, with only a few notable exceptions. However, the effects represented by OIRF may differ between the time-invariant and time-varying models and at various time points in the time-varying GVAR models. Additionally, we form the OIRF to shocks from multiple variables by adding several responses of the same variable to the shocks from a variety of other economic variables. As a result, we can determine the change in a particular variable following shocks from more than one variable. The output suggests that a single variable often dominates the OIRF to multivariate shocks because of the stark disparity in magnitude between the OIRF to shocks from different variables.

More importantly, the findings of the impulse response can explain a wide range of real-world economic phenomena. In particular, these findings can shed light on the shifts in the relationship between various economies,  internally and the global economy, in response to financial crises. For example, in the time-invariant GVAR model, we already know that in Europe, a rise in GDP brings a fall in HUR (the third subgraph of the second row of Figure \ref{fig3}(a)).With the assistance of Okun's Law, it is conceivable to see the connection between GDP and HUR theoretically. The principles established by this law state that for every established one percent gain in GDP, there is a matching two percent increase in employment. The basis for this law is relatively straightforward, which states that the principles of demand and supply determine GDP levels. Hence an increase in demand will increase GDP. Such an increase in demand necessitates an equal rise in productivity and employment to meet the demand. Regarding the time-variant model, we have revealed that when the global COVID-19 epidemic exceeds 10 million confirmed cases, HUR in Europe falls when international oil prices rise (the second subgraph of the last row of Figure \ref{fig3}(b)). At the same time, it tends to increase on general occasions (the second subgraph of the last row of Figure \ref{fig3}(a)). Oil price increases typically result in higher pricing for services like production expenses, energy costs, and petroleum commodities. Therefore, the decrease in productivity will consequently impact interest rates, inflation rates, investments, product selling prices, consumption levels, HUR, and actual wage rates  \citep{kisswani2017effect}. That is, when oil prices rise, the increase will cause a reduction in employment. However, this relationship flipped at the time of the epidemic, and such change, in fact, also occurred in the two remaining economic crises (the second subgraph of the last row of Figure \ref{fig3}(c)(d)). Most crucially, traditional impulse response analysis is incapable of capturing this phenomenon because such significant economic upheavals should only be viewed momentarily and have a transient impact. Finally, in terms of the OIRF to shocks from more than one variable, the effect of crude oil prices on the European economy is considerably more significant than the impact itself during specific economic crises (Figure \ref{fig:OIRF_n_to_1}(c2)(d2)). It may be caused by the fact that Europe is highly reliant on oil for its economic development and that a global financial crisis can cause shortages of oil supplies and fluctuating oil prices. Thus, international crude oil prices have significantly impacted the European economy amid economic turmoil.

Our approach primarily provides an idea and a meant for out-of-sample prediction of time-varying GVAR, and designs a time-varying impulse response strategy. However, one caveat of our proposed model is that there is no strict theoretical guarantee, and the relatively strong assumptions about the stationary of the data need to be satisfied. Therefore, there is no guarantee that all pulses will converge (some pulses may diverge). Also, the LASSO-type technique does not provide optimal predictions for all economic variables. However, we provide a collection of verification screening strategies on the test set and the corresponding codes (\url{https://github.com/kannyjyk/TVP-GVAR-ML}). Finally, specific machine learning approaches make it possible to make more accurate predictions regarding the possible temporal structures.

\clearpage
\newpage

\appendix
\section{The proof of the asymptotic distribution of OIRF}

Suppose
$$
\sqrt{T}
\left[
\begin{array}{c}
	\hat{\bm{\alpha}} - \bm{\alpha} \\
	\hat{\bm{\sigma}} - \bm{\sigma} \\
\end{array}
\right]
\stackrel{d}{\rightarrow}
\mathcal{N}
\left(
0,
\left[
\begin{array}{cc}
	\bf{\Sigma}_{\hat{\bm{\alpha}}} & 0 \\
	0 & \bf{\Sigma}_{\hat{\bm{\sigma}}} \\
\end{array}
\right]
\right)
$$

Then according to the multivariate Delta method we know that
$$
\sqrt{T}\ \text{vec}(\widehat{\mathbf{OImp}}(n) - \mathbf{OImp}(n)) \stackrel{d}{\rightarrow} \mathcal{N}(0,\bm{C}_n\bm{\Sigma}_{\widehat{\bm{\alpha}}} \bm{C}'_n + \bm{\bar{C}}_n\bm{\Sigma}_{\widehat{\bm{\sigma}}}\bar{\bm{C}}'_n)
$$

Therefore, with the assistance of the product rules for vector differentiation, we have
$$
C_i = \frac{\partial\text{vec}(\textbf{OImp}(n))}{\partial\bm{\alpha}'} = \frac{\partial\text{vec}(\textbf{B}_n\bm{P}_\varepsilon)} {\partial\bm{\alpha}'} = (P'\otimes I_K) \frac{\partial\text{vec}(\textbf{B}_n)} {\partial\bm{\alpha}'}
$$

and
$$
\bar{C}_i = \frac{\partial\text{vec}(\textbf{OImp}(n))}{\partial\bm{\sigma}'} = (I_K\otimes \textbf{B}_n) \frac{\partial\text{vec}(\bm{P}_\varepsilon)} {\partial\bm{\sigma}'}
$$

where
$$
\frac{\partial\text{vec}(\bm{P}_\varepsilon)}{\partial\bm{\sigma}'} = \bm{L}_K' \frac{\partial\text{vech}(\bm{P}_\varepsilon)}{\partial\bm{\sigma}'} = \bm{H}
$$

\clearpage
\section{Declaration of competing interest}
None

\section{Acknowledgements}
The authors would like to thank the Organisation for Economic Co-operation and Development (OECD) for collecting the economic data and their efforts. 





\bibliographystyle{Chicago}
\bibliography{mybibtex}
\end{document}